\documentclass[preprint]{iucr}              % DO NOT DELETE THIS LINE

\usepackage{siunitx}
\usepackage{physics}
\usepackage{nicefrac}
\usepackage{xspace}
\usepackage{booktabs}
\usepackage[table]{xcolor}
\usepackage{multirow}
\usepackage{makecell}
\usepackage{enumitem}

\DeclareSIUnit[number-unit-product={}]{\degTTh}{\SIUnitSymbolDegree\ensuremath{~2\mathrm{\theta}}}
\DeclareSIUnit[number-unit-product={}]{\TTh}{\ensuremath{2\mathrm{\theta}}}
\DeclareSIUnit[number-unit-product={}]{\step}{step}
\DeclareSIUnit[number-unit-product={}]{\wtpercent}{~wt\%}

\newcommand{\Rexp}{$R_\mathrm{exp}$\xspace}
\newcommand{\Rp}{$R_\mathrm{p}$\xspace}
\newcommand{\Rwp}{$R_\mathrm{wp}$\xspace}
\newcommand{\RBragg}{$R_\mathrm{Bragg}$\xspace}

     \journalcode{J}              % Indicate the journal to which submitted
\begin{document}                  % DO NOT DELETE THIS LINE

     %-------------------------------------------------------------------------
     % The introductory (header) part of the paper
     %-------------------------------------------------------------------------

     % The title of the paper. Use \shorttitle to indicate an abbreviated title
     % for use in running heads (you will need to uncomment it).

\title{The effect of data quality and model parameters on the quantitative phase analysis of X-ray diffraction data by the Rietveld method}
\shorttitle{Quantitative phase analysis by Rietveld}

\cauthor[a]{Matthew R.}{Rowles}{matthew.rowles@curtin.edu.au}

\aff[a]{John de Laeter Centre, Curtin University, Perth, WA \country{Australia}}

\maketitle                        % DO NOT DELETE THIS LINE

\begin{synopsis}
The step size, angular range, and intensity of X-ray powder diffraction data and the number and type of refinable parameters has been examined with respect to their effect on quantitative phase analysis by the Rietveld method.
\end{synopsis}

\begin{abstract}
The quality of X-ray powder diffraction data and the number and type of refinable parameters have been examined with respect to their effect on quantitative phase analysis (QPA) by the Rietveld method using data collected from two samples from the QPA round robin [Madsen \textit{et al. J. Appl. Cryst.} (2001), \textbf{34}, 409--26]. From these analyses of these best-case-scenario specimens, a series of recommendations for minimum standards of data collection and analysis are proposed. It is hoped that these will aid new QPA-by-Rietveld users in their analyses.
\end{abstract}

     %-------------------------------------------------------------------------
     % The main body of the paper
     %-------------------------------------------------------------------------
     % Now enter the text of the document in multiple \section's, \subsection's
     % and \subsubsection's as required.

\section{Introduction}

Quantitative phase analysis (QPA) by powder diffraction is one of the only truly phase-sensitive methods of analysis, as it can distiguish and quantify phases based on their unique crystal structures \cite{Madsen-VolH-2019}. Within the QPA of diffraction data, analyses can be divided into two separate groups: (i) single-peak methods, and (ii) whole-pattern methods. In single-peak methods, the intensity of a single peak, or group of peaks, is taken as representing the amount of that particular phase present in the specimen. These measurements can be biased due to peak overlap, or preferred orientation. In whole-pattern methods, diffraction data over a wide range is compared between unknowns and standards, where the standard data has been measured from pure specimens, or calculated from crystal structure information.

Of the whole-pattern methods, the Rietveld method \cite{Rietveld-JAC-1969,Loopstra-ACB-1969,Toby-VolH-2019} is widely used for QPA \cite{Madsen-JAC-2001,Scarlett-JAC-2002}, microstructural characterisation \cite{Balzar-JAC-2004, Scardi-2004}, and structure determination and refinement \cite{Hill-JAC-1992,LeBail-PD-2009,Peterson-AM-2011}. In this method, a diffraction pattern is calculated point by point \cite{Madsen-2013,Dinnebier-2019}, using

\begin{align} \begin{split}
I_{(hkl),\alpha} &= \left[ \frac{I_0 \lambda^3}{32\pi R} \frac{e^4}{m_e^2 c^4} \right] \times \left [  \frac{m_{(hkl)}}{2V_\alpha^2} \left | F_{(hkl),\alpha} \right|^2 \times \right. \\
&\left. LP \times \exp \left (-2 B_\alpha \left\{ \frac{\sin\theta}{\lambda} \right\}^2 \right) \right ] \times \left[ \frac{W_\alpha}{\rho_\alpha \mu_m^*} \right]
\label{eqn:Rietveld}
\end{split}\end{align}
where subscripts $hkl$ and $\alpha$ represent that from a specific reflection and/or phase, respectively. $I$ is the reflection intensity, $I_0$ is the incident beam intensity, $R$ is the specimen-detector distance, $\lambda$ is the X-ray wavelength, $\left. e^4 \right / \left(m_e^2 c^4\right)$ is the classical electron radius, $m$ is the reflection multiplicity, $V$ is the unit cell volume, $F$ is the structure factor, $LP$ is the Lorentz-polarisation correction, $B$ is the overall atomic displacement parameter, $\theta$ is the Bragg angle, $\rho$ is the phase density, $\mu_m^*$ is the mass attenuation coefficient of the entire specimen, and $W$ is the weight fraction. This equation is augmented by a peak shape function representing microstructural and instrumental parameters, such as crystallite size and beam divergence, respectively, which could be based on empirical functions \cite{Young-ANM-1989}, or on fundamental parameters \cite{Cheary-JAC-1992}. 

In a Rietveld refinement, the model parameters are optimised by the method of nonlinear least-squares in order to minimise the distance between the model and the data, according to some metric \cite{Young-1995,Toby-PD-2012}. \citeasnoun{McCusker-JAC-1999} has published guidelines, based on crystal structure analysis, for Rietveld refinement. The order in which to refine parameters in a model is discussed by \citeasnoun{Toby-VolH-2019} -- see also Appendix A -- which has the guiding principle of ``refine parameters which will make the biggest difference first''. During refinement, the  correlation matrix should be regularly inspected to determine which parameters affect other parameters of interest.

As the relative intensities of the peaks are set by the crystal structure and instrumental parameters, the overall absolute intensity can be represented by a scale factor, $s$, allowing Equation \ref{eqn:Rietveld} to be simplfied as
\begin{equation}
s_\alpha = K\frac{1}{V_\alpha^2} \frac{W_\alpha}{\rho_\alpha}\frac{1}{2 \mu_m^*}
\label{eqn:sf}
\end{equation}
where $K$ is a constant dependent only on the instrumental conditions \cite{OConnor-PD-1988}. As $\rho_\alpha = \left . M_\alpha\right / V_\alpha$\footnote{Note that \SI{1}{\dalton\per\cubic\angstrom} = \SI{1.660529}{\gram\per\cubic\cm} }, where $M_\alpha$ is the mass of the unit cell\footnote{In some texts, the mass of the unit cell is given as $ZM$, where $Z$ is the number of formula units, and $M$ is the mass of one formula unit.}, we can substitute and rearrange Equation~\ref{eqn:sf} to give
\begin{equation}
W_\alpha = \frac{(sMV)_\alpha \mu_m^*}{K'}
\end{equation}
which allows for quantification by the so-called the external standard method \cite{OConnor-PD-1988}. From this, \citeasnoun{Hill-JAC-1987b} and \citeasnoun{Bish-JAC-1988} applied the constraint of $\sum_{k=1}^n W_k = 1$ \cite{Chung-JAC-1974a,Chung-JAC-1974b} to give the equation most widely used in most Rietveld-based QPA:
\begin{equation}
W_\alpha = \frac{(sMV)_\alpha}{\sum_k (sMV)_k}
\end{equation}
In this equation, the scale factors, acting as a proxy for the measured intensities, are calibrated by the phase constant $MV$. Note that if the unit cell parameters or site occupancies are refined, this constant is dynamically updated. It is important to note that the weight fraction returned from this method is in relative, not absolute, terms. If absolute quantification is needed, then an internal \cite{Westphal-JAC-2012} or external standard \cite{OConnor-PD-1988} approach should be used; this is especially true in analysis of \textit{in situ} data (\S3.9.7)\cite{Madsen-VolH-2019}. If no, or only partial, structural information is available, then an alternative process must be sought, such as calibrated hkl files \cite{Taylor-PD-1992,Scarlett-PD-2006}, or the Direct-Derivation Method \cite{Toraya-JAC-2016}.

QPA by the Rietveld method has been the focus of many round-robins studying Portland cements \cite{Leon-Reina-JAC-2009}, minerals \cite{Raven-2017, Madsen-JAC-2001,Scarlett-JAC-2002}, clays \cite{Raven-CCM-2017}, ceramics \cite{Toraya-JAC-1999}, and pharmaceuticals \cite{Fawcett-PD-2012, Scarlett-JAC-2002}, as well as studies investigating the effect of radiation type \cite{Leon-Reina-JAC-2016}, and studying the outcomes of round-robins \cite{Peplinski-MSF-2004,Whitfield-PD-2016}. There have only been a few studies looking at the effect of data quality on QPA results \cite{Madsen-2013,Uvarov-JAC-2019}, and structural refinement results \cite{Hill-JAC-1984,Hill-JAC-1986,Hill-PD-1987}. The ability to assess amorphous content has been studied \cite{Madsen-ZC-2011,Gualtieri-JAC-2014}, and minimum reporting guidelines have been proposed  \cite{Gualtieri-PdM-2019}.

QPA is important in many areas, such as \textit{in situ} and \textit{operando} experimentation \cite{Jorgensen-JPCC-2020,Brant-JPS-2016}, quality control/assurance in syntheses, and process monitoring in on-site quality control laboratories \cite{Scarlett-PD-2001}. When data is being collected solely for QPA, there is no point in collecting unnecessary, and therefore time-consuming, data which does not increase the accuracy of the phase abundance. This necessarily leads to the questions ``What are the properties of the data, in terms of measured intensity, \si{\TTh} range, and step size, required to support QPA analysis?'', and ``Which parameters should I optimise in a Rietveld refinement?''.

To this end, diffraction data were collected from two mixtures of well-characterised minerals \cite{Madsen-JAC-2001} with varying collection times to give varying measured intensities. Many automated Rietveld refinements \cite{Coelho-JAC-2018a} were conducted to assess the impact of  \si{\TTh} range and the number and type of model parameters on the QPA results. From these data and refinement results, minimum criteria for the robust collection and analysis of diffraction data for quantification by the Rietveld method are given.

\section{Experimental}

\subsection{Diffraction data collection and reduction}

The two samples considered in this study were taken from the Sample 1 suite from the IUCr Commission on Powder Diffraction round robin on quantitative phase analysis \cite{Madsen-JAC-2001}. The sample suite consisted of a three-phase mixture of varying proportions of corundum ($\mathrm{Al_2O_3}$), fluorite ($\mathrm{CaF_2}$), and zincite ($\mathrm{ZnO}$). These phases were originally chosen to give minimal peak overlap, a good distribution of peaks with diffraction angle, and to provide little microabsorption constrast. Original specimens of samples 1a and 1e were procured and used as-received. These two samples represent mixtures whose diffraction patterns show the most even (1e) and most disparate (1a) distributions in intensity between the three phases. 

The two specimens were front-loaded into a standard specimen holder, and multiple diffraction patterns of each sample collected with the instrument conditions outlined in Table~\ref{table:inst_cond} with a single loading of the specimen holder. The data collection time and tube current were manipulated, as given in Table~S1, such that the maximum intensity of each pattern ranged between approximately 100 and \num{100000} counts above background. 

% S1 == \ref{table:collection}

%%%%%%%%%%%%%%%%
\begin{table}
\caption{Summary of instrument conditions for data collection.}
\label{table:inst_cond}
\begin{center}
\begin{tabular}{lc}      % Alignment for each cell: l=left, c=center, r=right
Instrument							&	Bruker D8 Advance \\
Incident radiation						&	Cu K$\alpha$  \\
Tube voltage (\si{\kilo\volt})                 &    40 \\	
Divergence (\si{\degree})				&	0.3 \\
Incident Soller slit (\si{\degree})		&	2.5 \\
Diffracted Soller slit (\si{\degree})		&	2.5 \\
Diffracted beam filter 				      &	Ni \\
Detector type                           		&	LynxEye \\
Detector opening    (\si{\degree})		&	2.796 \\
Detector channels                          		&	177 \\
Instrument radius (\si{\milli\metre})	&	250 \\
Data range (\si{\degTTh})				&	\numrange{15}{150}
\end{tabular}
\end{center}
\end{table}
%%%%%%%%%%%%%%%%

\begin{figure}
\includegraphics{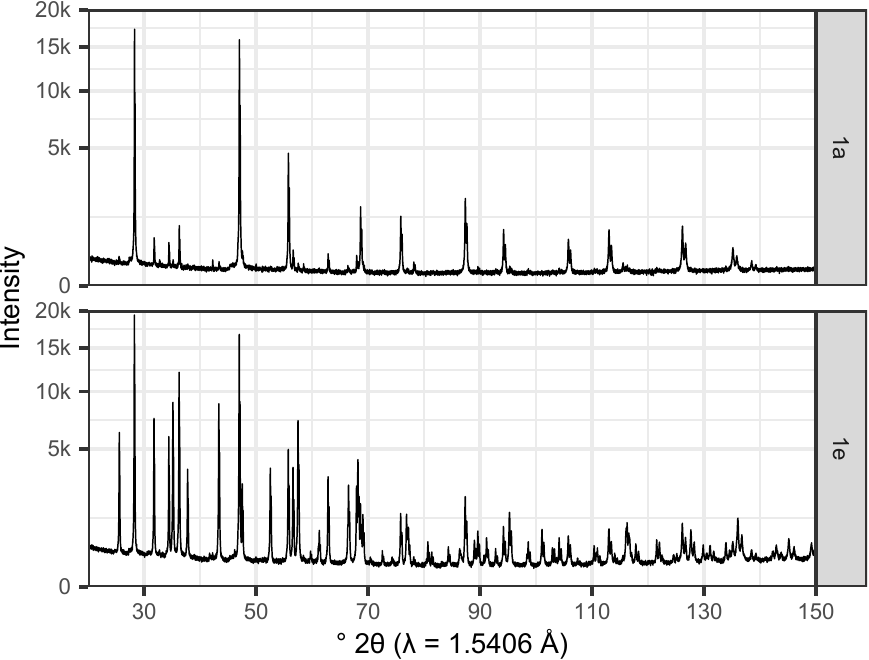}
\caption{Diffraction data for samples 1a and 1e collected with a nominal step size and maximum intensity of \SI{0.01}{\degTTh} and \num{20000} counts, respectively.}
\label{fig:diffdata}
\end{figure}

From these collected patterns, 13 diffraction patterns for each sample with nominal maximum intensities of 100 to \num{1000000} counts were calculated by summing up the requisite number of lower intensity patterns. Example diffraction patterns of samples 1a and 1e with a maximum intensity of \num{20000} counts are given in Figure~\ref{fig:diffdata}. To give diffraction patterns of different step sizes, data points were dropped from the original diffraction patterns in order to give eight different nominal step sizes logarithmically spaced from 0.01 to \SI{0.32}{\degTTh}. An example of how this affects the peaks is given in Figure~S1. Table~S2 gives the nominal and actual step sizes and maximum intensities. 

% S1 == \ref{fig:stepsize}
% S2 == \ref{table:nomact}

%%%%
%%%%

\subsection{Model development}

For maximum applicability, the model refinement process was automated to ensure repeatability and uniformity of refinement between specimens and diffraction patterns. Four different refinement model types were used:
\begin{enumerate}
\item\label{itm:first}     Refining only background, scale factors, and specimen displacement;
\item\label{itm:second} Refining~\ref{itm:first} and unit cell parameters;
\item\label{itm:third}    Refining~\ref{itm:second} and crystallite size, microstrain, and mixture packing density; and
\item\label{itm:forth}    Refining~\ref{itm:third}  and atomic displacement parameters (ADPs).
\end{enumerate}

These refinement types mimic approaches that may be used in various applications. For example, in a process-control application, variables such as unit cell parameters and crystallite size may be fixed at previously determined values, and only background and scale factors allowed to refine in any analysis. In a standard quality control analysis, additional parameters can be refined, for example, unit cell parameters and crystallite size and microstrain. In an \textit{in situ} experiment involving heating and reactions, unit cell parameters, crystallite size and microstrain, and ADPs would all need to be refined in order to properly model the evolution of the phases present in the system. It should be noted that there is a correlation between ADPs and scale factors, and which will affect QPA \cite{Madsen-ZC-2011}. This may be mitigated, in an \textit{in situ} analysis, by refining the models in a parametric manner \cite{Stinton-JAC-2007}.

The refinement strategies outlined above were applied to all diffraction patterns over an angular range from a constant low angle limit of \SI{21}{\degTTh} to high-angle limits (HALs) of 40, 50, 70, 90, 110, 130 and \SI{150}{\degTTh}. A specimen transparency peak-shape aberration was introduced using a refineable packing density as a multiplier of the specimen linear absorption coefficient\footnote{As calculated from the crystalline density and mass absorption coefficient of each phase combined by the quantitative phase analysis of the model.} (\S 4.4)\cite{Cheary-JNIST-2004}. This peak-shape aberration was not present in the model until the packing density was refined. The contribution of crystallite size and microstrain to the peak shape was modelled by the double-Voigt method \cite{Balzar-AXA-1995}, where these contributions are a combination of Lorentzian and Gaussian peak shapes. Lorentzian crystallite size broadening was present from the first step, but the parameter was not refined until its turn. Microstrain broadening was not included until its respective parameters were refined. No preferred orientation or anisotropic peak broadening parameters were used, as an exploratory data analysis showed that none of these were required to explain the peak shapes or intensities.

The details of the parameter refinement order are given in Appendix A. In summary, parameters were refined in the following order:
\begin{enumerate}[label=\alph*]
  \setlength{\itemsep}{0pt}
  \setlength{\parskip}{0pt}
\item\label{itm:first_ref}       Fifth-order Chebyshev polynomial background and scale factors, with a fixed Lorentzian crystallite size;
\item\label{itm:second_ref}  Refining \ref{itm:first_ref} and unit cell parameters;
\item\label{itm:third_ref}      Refining \ref{itm:second_ref} and specimen displacement;
\item\label{itm:fourth_ref}    Refining \ref{itm:third_ref} and Lorentzian crystallite size;
\item\label{itm:fifth_ref}       Refining \ref{itm:fourth_ref} and packing density;
\item\label{itm:sixth_ref}      Refining \ref{itm:fifth_ref} and Gaussian microstrain;
\item\label{itm:seventh_ref}  Refining \ref{itm:sixth_ref} and Gaussian crystallite size and Lorentzian microstrain;
\item\label{itm:eighth_ref}    Refining \ref{itm:seventh_ref} and ADPs; and
\item\label{itm:ninth_ref}      Refining \ref{itm:eighth_ref} again.
\end{enumerate}

At each step, the parameters were refined to convergence\footnote{The refinement was halted when the change in \Rwp in consecutive iterations was less than \num{0.001}.}, parameter estimates saved, and then the refinement restarted at the next step using the converged estimates of the previous step as the starting point. Parameter starting values were either fixed, or chosen from a uniform random distribution, as outlined in Table~\ref{table:modelstart}. Background and specimen displacement started at zero, as no other sensible value exists. The starting values for scale factors encompass an order of magnitude above and below their optimum value. Crystallite size and microstrain starting values cover those values typically seen in many materials. Unit cell parameter starting values cover the range that would reliably automatically converge to the correct result.

%%%%%%%%%%%%
\begin{table}
\caption{Model starting parameter values and refinement limits. Corundum, fluorite, and zincite are abbreviated as cor, flu, and zin. Numbers in brackets indicate a number chosen at random, uniformly the two given values. Values were allowed to refine between the given limits, or without limit, as indicated by '--'. A single value preceeded by a ``!'' indicates it was fixed during refinement.}
\label{table:modelstart}
\begin{center}
\begin{tabular}{rrrccc}
\toprule
\multicolumn{3}{c}{\multirow{2}{*}{Parameter}}                        & \multirow{2}{*}{Value / range} & \multicolumn{2}{c}{Limits} \\
\multicolumn{3}{c}{}                                                  &                                & min          & max         \\
\midrule
\multicolumn{3}{r}{Chebyshev background}                                        & 0 0 0 0 0 0 0             &      --        &      --       \\
\multicolumn{3}{r}{Zero error (\si{\degTTh})}                                    & !0                              &               &          \\
\multicolumn{3}{r}{Specimen displacement (\si{\milli\m})}                   & 0                             & -0.5         & 0.5         \\
\multicolumn{3}{r}{Packing density}                                                   & (0.1, 0.7)                  & 0.001        & 10          \\
\midrule
\multirow{6}{*}{Scale factor}            
                                  & \multirow{2}{*}{cor}        & 1a  & (0.044, 4.4)$\: \times \: 10^{-3}$               & 0            &      --       \\
                                  &                                         & 1e  & (0.0057, 0.57)                     & 0            &      --       \\
                                  & \multirow{2}{*}{flu}         & 1a  & (0.0073,0.73)                  & 0            &     --        \\
                                  &                                         & 1e  & (0.0086, 0.86)                 & 0            &   --          \\
                                  & \multirow{2}{*}{zin}        & 1a  & (0.0018, 0.18)                 & 0            &      --       \\
                                  &                                         & 1e  & (0.028, 2.8)                   & 0            &    --         \\
\midrule
\multirow{2}{*}{\makecell{Crystallite size \\ (\si{\nano\m})}} & \multicolumn{2}{r}{Lorentzian}    & (50, 500)                      & 20           & \num{10000}       \\
                                  & \multicolumn{2}{r}{Gaussian}      & !\num{10000}                          &              &             \\
\multirow{4}{*}{Microstrain}      & \multirow{3}{*}{Lorentzian} & cor & (0.01, 0.1)                    & 0.0001       & 6           \\
                                  &                             & flu & (0.01, 0.1)                    & 0.0001       & 6           \\
                                  &                             & zin & !0.0001                         &              &             \\
                                  & Gaussian                    &     & !0.0001                         &              &             \\
\midrule
\multirow{5}{*}{\makecell{Unit cell \\ parameter (\si{\angstrom})}}         & \multirow{2}{*}{cor}        & a   & (4.75, 4.77)                   & 4.7          & 4.8         \\
                                  &                             & c   & (12.94, 13.04)                 & 12.93        & 13.05       \\
                                  & flu                         & a   & (5.44, 5.48)                   & 5.4          & 5.5         \\
                                  & \multirow{2}{*}{zin}        & a   & (3.24, 3.26)                   & 3.2          & 3.3         \\
                                  &                             & c   & (5.19, 5.23)                   & 5.15         & 5.25        \\
\midrule
\multicolumn{3}{r}{\makecell{Isotropic atomic \\ displacement parameter (\si{\angstrom\squared})}}               & (0, 1)                         & -10          & 10          \\
\bottomrule
\end{tabular}
\end{center}
\end{table}
%%%%%%%%%%%%%%%%

The models were refined starting from refinement type 4, and proceeding to 1. When moving to the next refinement type, the value of the parameters which were newly fixed were taken as the average of the estimates of those parameters from the models refined for sample 1e, step size = \SI{0.01}{\degTTh}, HAL = \SI{150}{\degTTh}, and maximum intensity $\ge$ \num{20000}. These values, given in Table~\ref{table:modelfinish}, were used for both 1a and 1e, as the distribution of intensities in 1e is better than those in 1a.

A second set of refinements were also undertaken with the zero error refining in conjunction with the specimen displacement; they were otherwise identical. The equivalent figures from these data are presented in the supplementary information.

%%%%%%%%%%%%
\begin{table}
\caption{Parameter values used when, according to the refinement type, a given parameter was fixed.}
\label{table:modelfinish}
\begin{center}
\begin{tabular}{rrrS[table-format=3.6]}
\toprule
\multicolumn{3}{c}{Parameter}                                    & {Value}    \\
\midrule
\multicolumn{3}{r}{Packing density}                              & 0.172    \\
\midrule
\multirow{3}{*}{\makecell{Crystallite size \\ Lorentzian (\si{\nano\m})}}    
                                     & \multicolumn{2}{r}{cor}   & 311 \\
                                     & \multicolumn{2}{r}{flu}   & 590      \\
                                     & \multicolumn{2}{r}{zin}   & 292      \\
\midrule
\multirow{2}{*}{\makecell{Microstrain \\ Lorentzian}}         
                                     & \multicolumn{2}{r}{cor}   & 0.0193   \\
                                     & \multicolumn{2}{r}{flu}   & 0.0415   \\
\midrule
\multirow{5}{*}{\makecell{Unit cell \\ parameter (\si{\angstrom})}} 
                                     & \multirow{2}{*}{cor} & a  &   4.759355 \\
                                     &                                  & c  & 12.992743 \\
                                     & flu                             & a  &  5.464412 \\
                                     & \multirow{2}{*}{zin} & a  &   3.249915 \\
                                     &                                  & c  &   5.206751  \\
\midrule
\multirow{6}{*}{\makecell{Isotropic atomic \\ displacement \\ parameter (\si{\angstrom\squared})}}   
                                     & \multirow{2}{*}{cor} & Al & 0.249    \\
                                     &                                  & O  & 0.193    \\
                                     & \multirow{2}{*}{flu} & Ca & 0.467    \\
                                     &                                  & F  & 0.692    \\
                                     & \multirow{2}{*}{zin} & Zn & 0.523    \\
                                     &                                  & O  & 0.321   \\
\bottomrule
\end{tabular}
\end{center}
\end{table}
%%%%%%%%%%%%

The refinements were conducted in TOPAS Academic v6 \cite{Coelho-JAC-2018a}, making use of the in-built macro language and command line capabilities for automation. The instrumental profile was modelled using fundamental parameters \cite{Cheary-JNIST-2004, Cheary-JAC-1992}, but as these parameters were fixed at known values, this approach would be equally valid for other empirically-defined models. By default, TOPAS minimises \Rwp based on Newton-Raphson non-linear least squares, with the Marquardt method \cite{Marquardt-JSIAM-1963,Coelho-JAC-2018b} included for stability. The convergence of the non-linear least squares process is aided by the bound constrained conjugate gradient method \cite{Coelho-JAC-2005}. The model was calculated only at the data points, using the condition \texttt{x\_calculation\_step = Yobs\_dx\_at(Xo);}. In TOPAS, the estimated standard deviations (esds) of each parameter estimate are calculated by singular value decomposition, and are given by the square root of the diagonal elements of the covariance matrix multiplied by the goodness-of-fit (GoF) -- see Equation~\ref{eqn:gof}. esds of derived parameter estimates take into account the correlation of the consituent parameters. For the purposes of this paper, the esds given by TOPAS have been divided by the GoF in order to retrieve the values from the covariance matrix \cite{Schwarzenbach-ACA-1989}, and any further reference to ``esd'' denotes a value taken as the square root of the appropriate value from the correlation matrix. 

Refined and derived parameter estimates, their esds, and figures-of-merit were written to file at the conclusion of each refinement. The derived parameters estimates include volume-weighted crystallite domain size (L\textsubscript{vol}) and strain (e\textsubscript{0}) \cite{Balzar-AXA-1995}, weight fraction using the \citeasnoun{Hill-JAC-1987b} approach, and the specimen’s linear attenuation coefficient. The figures-of-merit included \RBragg, \Rp, \Rwp, \Rexp, and GoF \cite{Young-1995}. The weighted Durbin-Watson ($d$) statistic was also calculated \cite{Hill-JAC-1987a}. 

These refinements were carried out 200 times for each set of conditions, diffraction pattern, and sample to obtain a distribution of estimates, resulting in 2 samples $\times$ 4 refinement types $\times$ 7 HALs $\times$ 8 step sizes $\times$ 13 maximum intensities $\times$ 200 repeats = \num{1164800} total refinements. This ordering provides a nomenclature with which to refer to specific refinements -- sample/refinement type/HAL/step size/maximum intensity, with missing values implicitly referring to all of those particular values. For example, 1e/4/70-110/$>$\num{30000} refers to sample 1e, refinement type 4, HAL = \SIrange{70}{110}{\degTTh}, all step sizes, and all maximum intensities greater than \num{30000}

\section{Results and discussion}

\subsection{Refinement output reduction and visualisation}

Data visualisation during model development was conducted using gnuplot \cite{Williams-2019}, which allowed for automated, script-driven figure production to allow for easy inspection of parameter stability during refinements. After all model refinements were completed, R \cite{R-2020} and RStudio \cite{RStudio-2019} were used to carry out further analysis. Visualisations were carried out using ggplot2 and associated libraries \cite{Wickham-2016, Akima-CRAN-2016,Campitelli-CRAN-2020,Campitelli-CRAN-2020b,Garnier-CRAN-2018,Pedersen-CRAN-2019,Schloerke-CRAN-2020,Slowikowski-CRAN-2020,Wickham-CRAN-2019}. 

%Additional calculated parameters include DDM quantification \cite{Toraya-JAC-2016}, Q  \cite{Hill-JAC-1987a,Theil-JASA-1961}, and relative errors. 

Within the 200 refinements conducted for each combination of sample, refinement type, HAL, step size, and maximum intensity, the estimates were summarised by their mean, standard deviation (sd), first and third quantiles, minimum, maximum, and standard uncertainty $\left( \mathrm{u} = \left. \sqrt{\sum \mathrm{esd}^2} \right / 200 \right )$, leaving \num{5824} results. These data are available in the supplementary information\footnote{Available at https://cloudstor.aarnet.edu.au/plus/s/j4V2SNKFJTX5uBH}. The original data is available online \cite{Rowles-data-2020b, Rowles-data-2020}. Some visualisations use a bisymmetric logarithmic scaling to allow for negative and zero values \cite{Webber-MST-2013}.

\subsection{Figures-of-merit}

One of the measure of “success” of a refinement are the figures-of-merit \Rwp and GoF, as well as the Durbin-Watson statistic, $d$ \cite{Hill-JAC-1987a,Young-1995,Toby-PD-2012}. These are given as
\begin{align}
%R_\mathrm{p} &= \frac{\sum\abs{y_m^\mathrm{obs} - y_m^\mathrm{calc}}}{\sum y_m^\mathrm{obs}} \\
R_\mathrm{wp} &= \sqrt{\frac{\sum w_m \qty(y_m^\mathrm{obs} - y_m^\mathrm{calc})^2 }{\sum w_m \qty(y_m^\mathrm{obs})^2}} \\
R_\mathrm{exp} &= \sqrt{\frac{M - P}{\sum w_m \qty(y_m^\mathrm{obs})^2}} \\
\mathrm{GoF} = \frac{R_\mathrm{wp}}{R_\mathrm{exp}} &= \sqrt{\frac{\sum w_m \qty(y_m^\mathrm{obs} - y_m^\mathrm{calc})^2 }{M-P}} \label{eqn:gof}\\
d &= \left . \qty[\sum_{m=2}^{M}\qty( \frac{\Delta_{m}}{\sigma_{m}} - \frac{\Delta_{m-1}}{\sigma_{m-1}} )^2] \right / 	\qty[\sum_{m=1}^M \qty( \frac{\Delta_{m}}{\sigma_{m} } )^2] \label{eqn:d}
\end{align}

\noindent where $y_m^\mathrm{obs}$  and $y_m^\mathrm{calc}$ are the $m^\mathrm{th}$ observed and calculated diffraction pattern intensities, $w_m=\sigma_m^{-2}$ is the weighting\footnote{If any data reduction or correction takes place, e.g. 2D to 1D azimuthal averaging \cite{Yang-JAC-2014}, the weighting of each data point must be correctly calculated and the correspoding uncertainty in intensity must reported in the final data file to allow analysis programs to apply the correct weighting to each datapoint.} of the $m^\mathrm{th}$  observed intensity, where, in this study, the uncertainty in the measured intensity is given as $\sigma_m=\sqrt{y_m^\mathrm{obs}}$. $M$ is the number of measured intensities in the diffraction pattern, $P$ is the number of parameters in the Rietveld model, and $\Delta_m=y_m^\mathrm{obs} - y_m^\mathrm{calc}$ .

\Rexp  is a constant for any given refinement, and lower values of \Rwp and GoF indicate “better” fits, subject to the physical and chemical reasonableness of the model \cite{Toby-PD-2012}. GoF values $< 1$ indicates either overfitting, or that the standard uncertainties in the measured intensities were overestimated. Values of $d$ significantly away from two show that there is serial correlation in the residuals due to the refinement model being inadequate, and that the calculated parameter esds are affected. 

\begin{figure}
\includegraphics{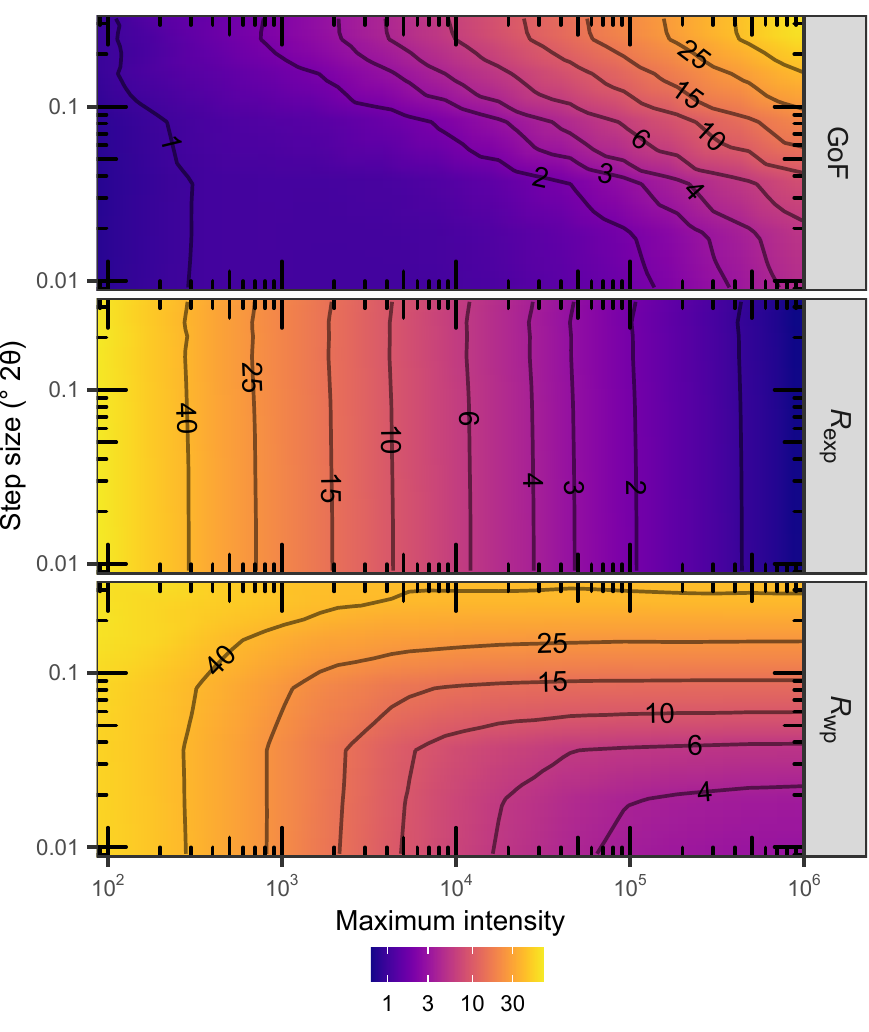}
\caption{The figures-of-merit, GoF, \Rexp, and \Rwp, for sample 1e/4/150. The trends evident in these plots are repeated throughout all the models. It can be seen that the desire for a low GoF and \Rwp are at odds with each other with respect to maximum intensity.}
\label{fig:fom}
\end{figure}

\begin{figure}
\includegraphics{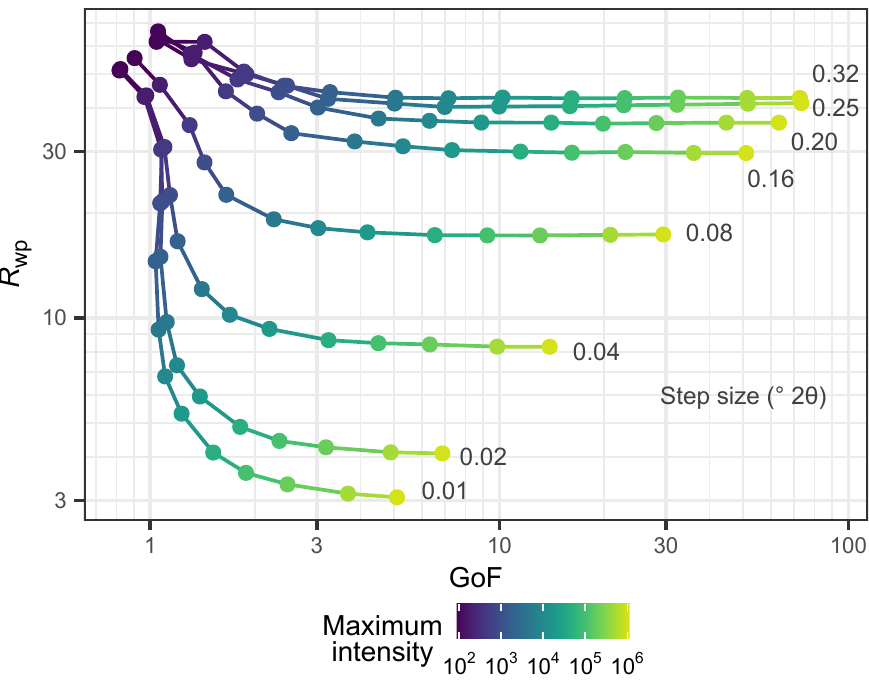}
\caption{Comparison of \Rwp and GoF for all sample 1e averaged over all refinement types and HALs, showing that step size is the best predictor low \Rwp and GoF, when coupled with a maximum intensity of $\sim$ \numrange{20}{50000} counts.}
\label{fig:gofrwp}
\end{figure}

Figure~\ref{fig:fom} shows how the GoF, \Rexp, and \Rwp values for sample 1e/4/150 change with step size and maximum intensity; the same trends are present for all samples. \Rexp is independent of step size, while \Rwp and GoF show competing behaviour to obtain the lowest value: \Rwp requires a high maximum intensity and GoF requires a low maximum intensity, with both requiring a small step size. The relationship between GoF and \Rwp is explored in Figure~\ref{fig:gofrwp}, where the \Rwp and GoF values from sample 1e are plotted averaged over all refinement types and HALs. This shows that the \Rwp and GoF can both be minimised by having a maximum intensity of $\sim$ \numrange{20}{50000} counts for step sizes in the range \SIrange{0.01}{0.04}{\degTTh}.

Note that the figures-of-merit should not solely be used to ascertain the correctness of a Rietveld model. The best way to determine the quality of a Rietveld model is by visual inspection of the observed and calculated patterns, noting any systematic variations in the residuals, and by ensuring that the final model is physically and chemically reasonable \cite{Toby-PD-2012}.

%%%%
%%%%

%%%%
%%%%

\subsection{Peak intensities}

The determination of QPA is predicated largely on the correct determination of peak intensities. In a Rietveld refinement, this is primarily controlled by the scale factor, but has secondary contributions \textit{via} peak width and shape parameters, such as crystallite size/microstrain and absorption, and through ADPs, which affect how the intensities in the model are distributed with angle. As scale factors are used directly in the QPA calculation, all of these parameters have a bearing on the calculated weight percentages.

The two main causes of systematic error in the determination of peak intensity are (i) microabsorption and (ii) preferred orientation. Microabsorption arises in a mixture containing both low- and highly-absorbing phases\footnote{Even in different sized particles of the same composition \cite{Whitfield-VolH-2019}.}. Highly-absorbing phases diffract primarily from their surface, and have their intensities, and hence QPA, underestimated, whereas low-absorbing phases are more likely to diffract from their bulk and over-contribute to the measured intensities. Microabsorption can only be known to be an issue through detailed prior knowledge of the constituent phases \cite{Scarlett-PD-2018}, and can sometimes be mitigated by grinding, choosing an X-ray wavelength which is equally absorbed by all phases, or by using neutron diffraction \cite{Scarlett-JAC-2002, Whitfield-PD-2016}. There exists the \citeasnoun{Brindley-PM-1945} correction, but this shouldn't be used unless the specimen consists of monodisperse, polycrystalline, spherical particles in a size range appropriate for the correction \cite{Cline-AXA-1982}. 

Preferred orientation arises when the perfect randomness of crystallite orientation is broken, for instance, with platy or needle-like particles. For many materials, the direction of this orientation can be taken as the basal plane or in a cleavage direction. For small amounts of preferred orientation, the changes in intensity can be accounted for by the March-Dollase \cite{Dollase-JAC-1986}, or spherical harmonics \cite{Jarvinen-JAC-1993} models; large deviations in intensity may necessitate variation in the specimen preparation or presentation \cite{Hillier-CM-1999}, or that alternate quantification procedures be undertaken.

Further changes in peak intensity can be caused by variations in site occupancy and atomic coordinates. Changes in site occupancy are common in many minerals, with partial and multiple occupancy depending on the mineral type, formation, and subsequent treatement. Refinement of these parameters in a mixture should be undertaken with care, and, if possible, the values should be constrained by known elemental composition, a previously determined relationship \cite{Fazey-CCM-1991}, or by a technique such as bond-valence sums \cite{Kaduk-PD-2009}. The refining of atomic coordinates moves the model from a quantitative phase analysis to a structure analysis. Great care should be taken in choosing which coordinates can be refined and if the resultant values describe a chemically sensible structure. 

Finally, peaks at low angles may have artificially low intensities due to beam overflow, where the equatorial divergence of the beam results in a beam footprint larger than the specimen at low angles. This can be overcome through using a fixed beam footprint length \textit{via} motorised slits, or collecting data with difference fixed slit settings over different \si{\TTh} ranges.

In the refinements reported herein, there was minimal microabsorption contrast present, no preferred orientation observed, fixed site occupancy and atomic coordinates, and there were no issues with beam overflow. Depending on the complexity of the system under investigation, any or all of these may need to be refined to obtain a correct result.

\subsubsection{Scale factors and peak area}

\begin{figure}
\includegraphics{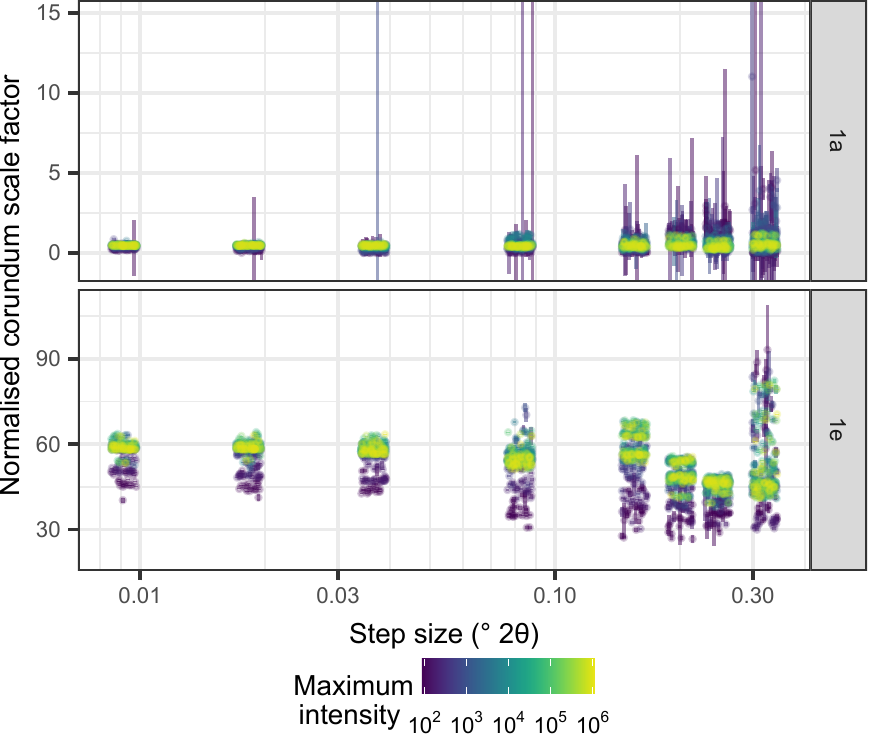}
\caption{Scale factors for corundum normalised by maximum intensity for samples 1a and 1e. Values move away from their ``correct'' position with decreasing maximum intensity and/or increasng step size. The error bars represent twice the standard uncertainty. The datapoints have been displaced slightly from their x-axis values for clarity.}
\label{fig:sf}
\end{figure}

The corundum scale factors for all refinement types and HALs are shown in in Figure~\ref{fig:sf}, normalised by maximum intensity. These data show that scale factors for the highest intensity data are stable with refinement type and HAL, and that as maximum intensity decreases and step size increases, the amount of spread in the scale factors increases. The spread in scale factors for sample 1a is limited by a lower possible value of zero, whereas the scale factors for 1e are able to spread in both directions. This is also realised by the relative uncertainty of the scale factor increasing much faster for sample 1a than 1e, showing that for specimens with minor/trace phases, small step sizes are necessary to yield more precise scale factors by capturing more of their intensity.

%%%%%
%%%%

\begin{figure}
\includegraphics{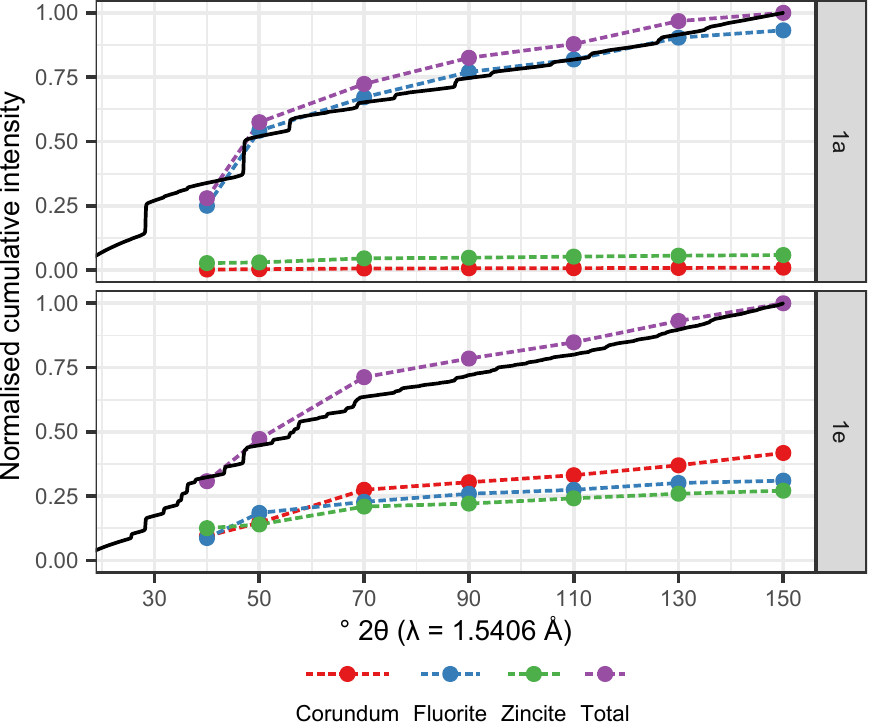}
\caption{The cumulative intensity of the diffraction data presented in Figure~\ref{fig:diffdata} and the numerical area of the individual phases and their sum normalised to the total diffracted intensity. In both samples 1a and 1e, it can be seen that intensities after \SI{70}{\degTTh} are evenly distibuted. The areas attributed to each phase change in relative distribution with low HAL values for both samples; after the intensities stabilise at \SI{70}{\degTTh}, their relative contributions remain constant.}
\label{fig:intdist}
\end{figure}

\begin{figure}
\includegraphics{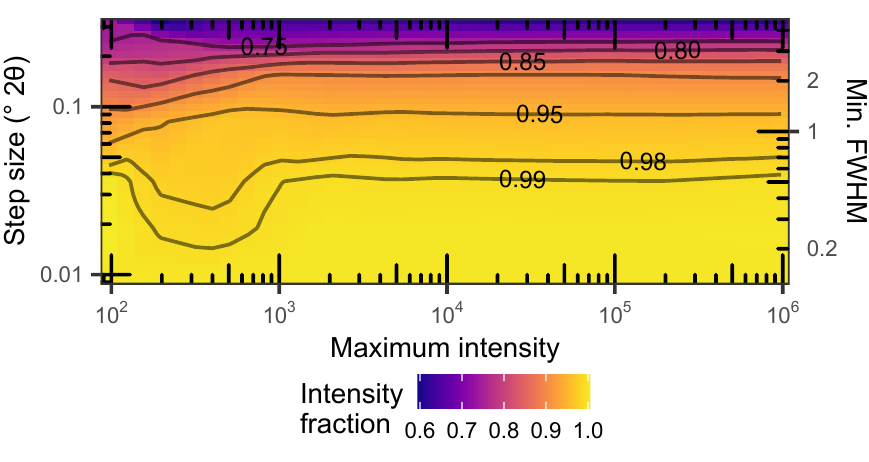}
\caption{Fraction of total intensity in sample 1e/3/150 as a ratio of the total intensity present in the \SI{0.01}{\degTTh} step size pattern. In order for QPA to be unaffected, intensity must be lost evenly from all phases; an unlikely event. The right axis gives the step size as a function of the average minimum peak full-width at half-maximum, as given in Figure~\ref{fig:fwhm}.}
\label{fig:numarea}
\end{figure}

The intensities present in sample 1e are evenly distributed for angles above \SI{70}{\degTTh} -- see Figure~\ref{fig:intdist} -- meaning that any increase in the HAL beyond this limit may stabilise other parameters in the refinement, but scale factors will remain largely unaffected, within the limits provided by the data quality and other refinable parameters. This may or may not apply to other mixtures and materials -- the cumulative intensity for sample 1a can be seen to follow the appearance of peaks of the major fluorite phase. The main factor which governs the measured intensities is the step size; as step size increases, the amount of ``lost'' intensity increases, as shown in Figure~\ref{fig:numarea}. In this figure, the integrated area of each model is shown, normalised to the area at a step size of \SI{0.01}{\degTTh} for each maximum intensity, showing how increasing the step size to $\sim$ \SI{0.1}{\degTTh} results in a loss of $\sim$ \SI{5}{\percent}. In order for this loss to not affect the QPA, all phases must lose the same amount of intensity at an identical stepsize. At step sizes approaching, and exceeding, the FWHM of the peaks, this is unlikely to occur, and so the accuracy of the QPA will be degraded.

%%%%
%%%%

%%%%
%%%%

\subsubsection{Crystallite size}

\begin{figure}
\includegraphics{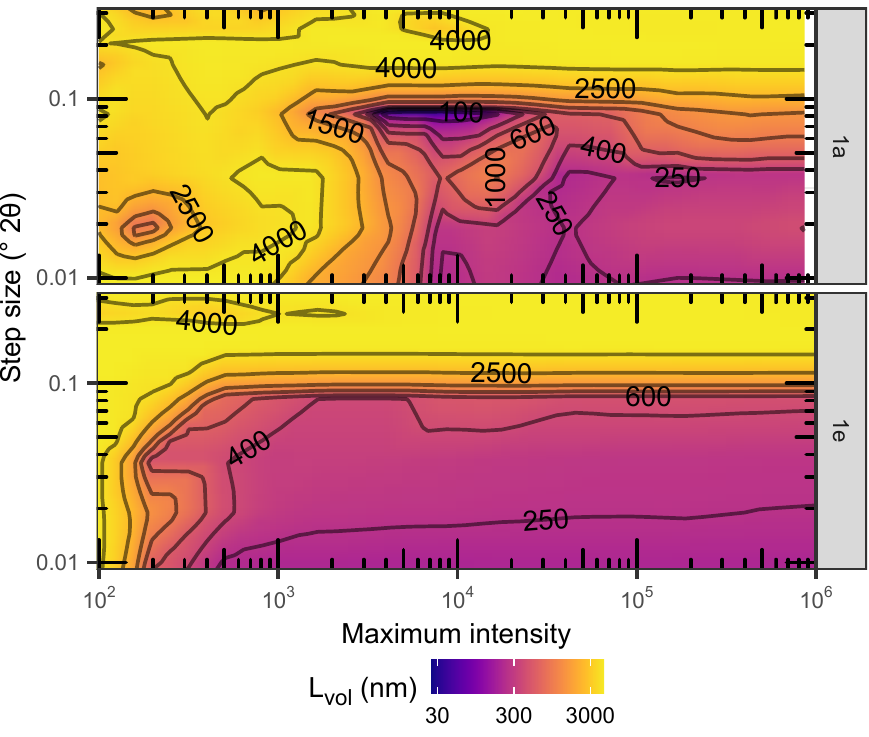}
\caption{Crystallite size for corundum in sample 3/150. It can be seen that they only agree for small step sizes and large maximum intensities, due to the ability to properly resolve the requisite peaks.}
\label{fig:cryssize}
\end{figure}

The crystallite size of corundum in sample 3/150 is shown in Figure~\ref{fig:cryssize}; the other phases follow the same trends. In both subfigures, there is a horizontal cut-off at a step size of approximately \SI{0.08}{\degTTh}, above which the crystallite size increases to physically unrealistic values. Additionally, there is a vertical cut-off to the left of which the crystallite size increases as the intensity of the peaks becomes small, allowing narrower peaks to fit in the noise of the diffraction data.

%%%%
%%%%

\begin{figure}
\includegraphics{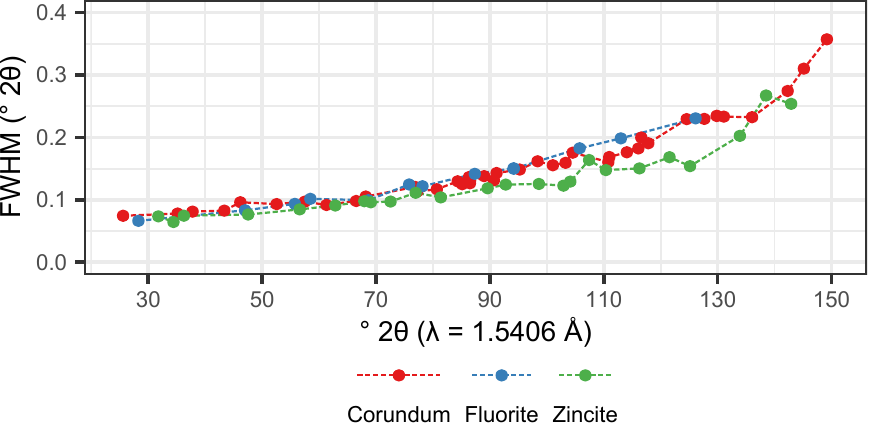}
\caption{The full-width at half-maximum for all peaks of all phases in sample 1e/0.01/\num{100000}.}
\label{fig:fwhm}
\end{figure}

The full-width at half-maximum (FWHM) of the peaks in the diffraction pattern from sample 1e/0.01/\num{100000} are shown in Figure~\ref{fig:fwhm}, and shows that the FWHM increases monotonically with diffraction angle, with the narrowest peaks having a FWHM $\approx$ \SI{0.07}{\degTTh}. This shows that biases in the crystallite sizes begin to appear when the step size in the data is the same size as the width of the first peak. It follows that a step size of at least half the minimum peak width is required to obtain realistic crystallite size results in a whole-pattern analysis context.

%%%%

\subsubsection{Absorption}

\begin{figure}
\includegraphics{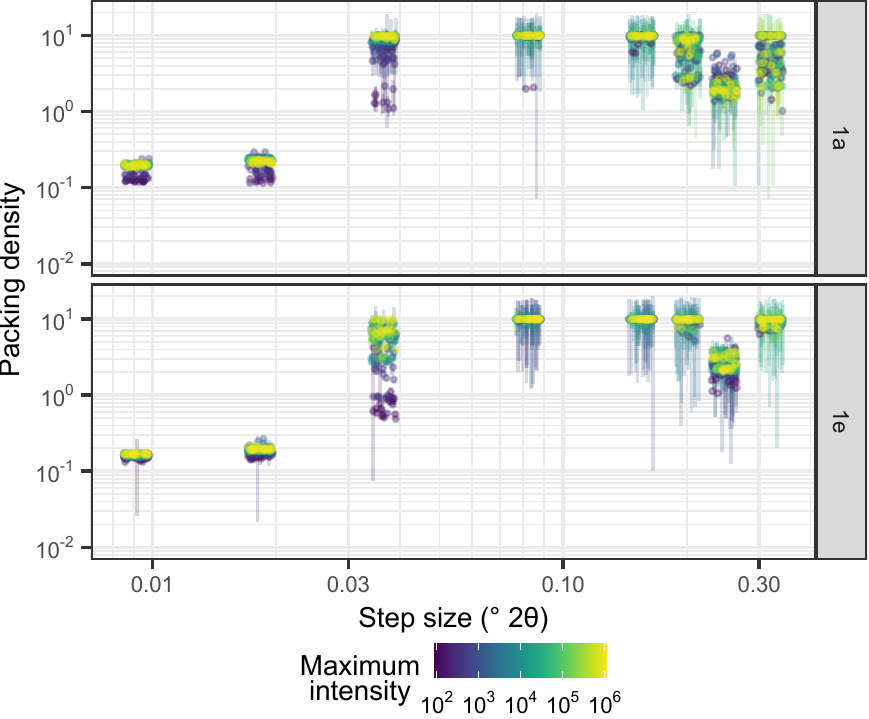}
\caption{Packing density for samples 1a and 1e for all intensities and HALs for refinement types 3 and 4. The refined value is highly dependent on step size, and only weakly dependent on maximum intensity. Error bars represent twice the standard uncertainty. The datapoints have been displaced slightly from their x-axis values for clarity.}
\label{fig:pd}
\end{figure}

The effect of absorption was introduced into the model through the form of an exponential peak convolution affecting the peak shape \cite{Cheary-JNIST-2004}. The key parameter for this convolution is the linear absorption coefficient of the specimen. In the model in this study, the theoretical linear absorption coefficient was calculated at each iteration from the elemental composition and calculated weight percentage of each phase. This value was modified by a packing density to scale the theoretical value to that actually exhibited by the specimen, where a value of 1 denotes a \SI{100}{\percent} dense material, and values in the range \numrange{0.1}{0.4} are typical. This packing density is shown in Figure~\ref{fig:pd} for all intensities and HALs for both samples.  With step sizes $\ge$ \SI{0.04}{\degTTh}, the refined packing density estimates quickly exceed physically reasonable values, indicating that the exponential broadening of the peaks due to absorption is no longer able to be distinguished, and the high values reflect that the model is unable to support any such additional broadening. It follows that a step size of at least a quarter of the minimum peak width is required to obtain realistic absorption values in the context of whole-pattern analysis, similar to that stated by \citeasnoun{McCusker-JAC-1999}.

%%%%
%%%%

\subsubsection{Atomic displacement parameters}

\begin{figure}
\includegraphics{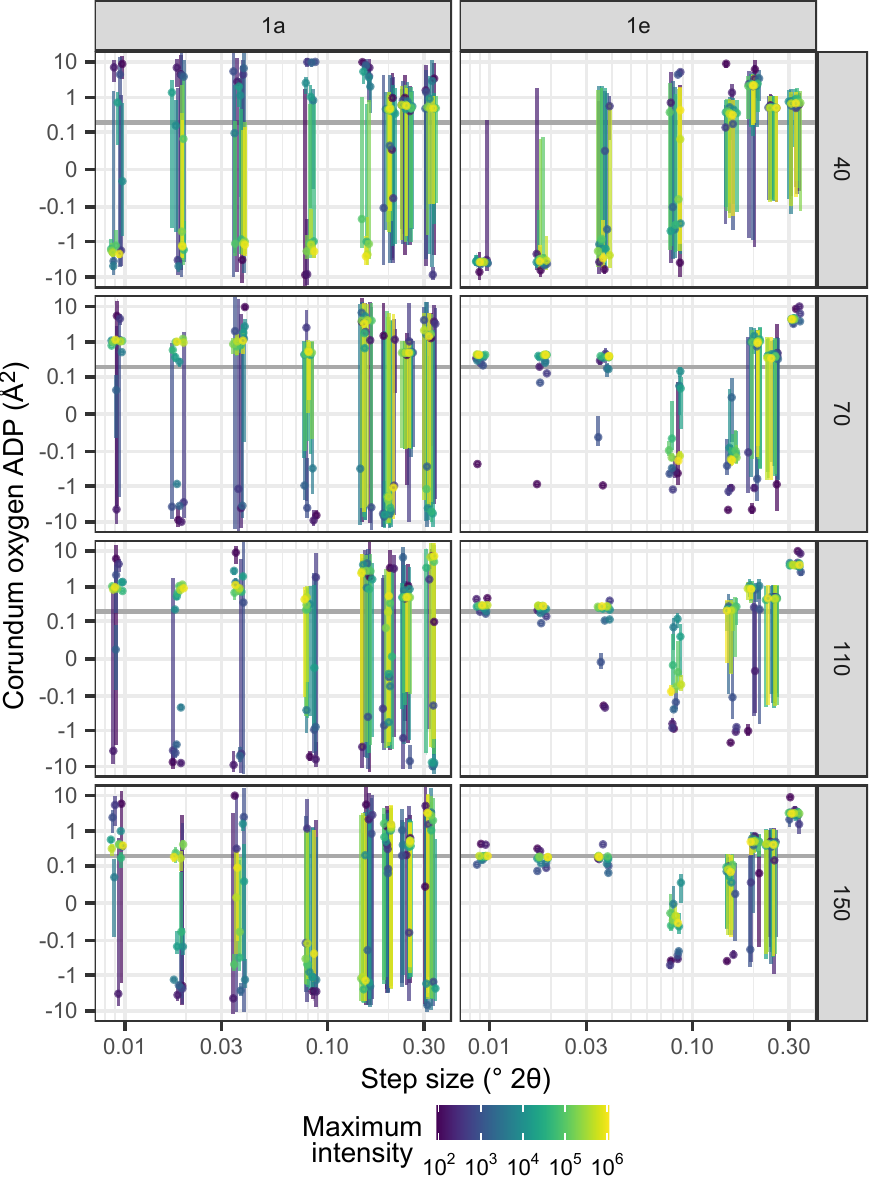}
\caption{ADP estimates for oxygen in corundum for  for samples 1a and 1e, refinement type 4, and the given HALs. Realistic values are only obtained for small step size, large HAL, and high maximum intensity data. The horizontal gray line represents the value given in Table~\ref{table:modelfinish}. The error bars represent twice the standard deviation, as the standard uncertainty is only of significance for low intensity, small HAL, large stepsize patterns. Please note that the vertial axis is logarithmic. The datapoints have been displaced slightly from their x-axis values for clarity.}
\label{fig:tp}
\end{figure}

Isotropic ADPs were refined for all atoms in all phases in refinement type 4. The parameter estimates for the oxygen ADP in corundum are shown in Figure~\ref{fig:tp}. It is clear that refinement of the ADPs for corundum in sample 1a, where the concentration is only \SI{1.15}{\wtpercent}, is not supported by the data; the intensities available in the data do not allow for the refinement of a physically realistic value with any combination of step size or HAL. For sample 1e, ADP refinment is not supported at any HAL for step sizes $>$ \SI{0.04}{\degTTh} due to the ``lost'' intensity corrupting the fine intensity detail required for such a refinement.

%%%%
%%%%

\begin{figure}
\includegraphics{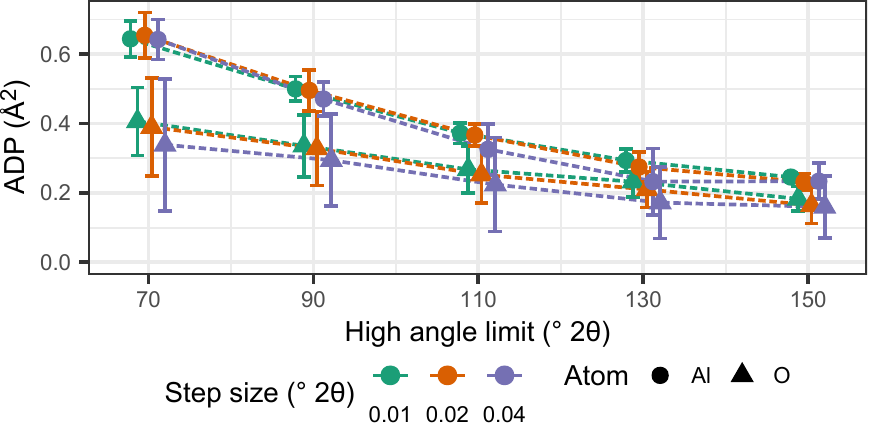}
\caption{ADPs for corundum in samples 1e/$>$~\num{1000}. Error bars represent twice the standard deviation of estimates averaged over all intensities. The datapoints have been displaced slightly from their x-axis values for clarity.}
\label{fig:tp2}
\end{figure}

This fine detail of the ADPs for corundum can be seen in Figure~\ref{fig:tp2}, where the evolution of the ADP estimates for Al and O with step size, intensity, and HAL is clear. The figure shows that the estimates reduce in scatter with smaller step size, and also continue to decrease with increasing HAL, leading to values of \SIrange[range-phrase = \,and\,]{0.248(14)}{0.18(4)}{\square\angstrom} for Al and O, respectively. These estimates differ from the single-crystal values of \SIrange[range-phrase = \,and\,]{0.19}{0.22}{\square\angstrom} \cite{Maslen-ACB-1993}, meaning that ADPs in mixtures should not be refined due to overlapping peaks\footnote{Temperature-dependent \textit{in situ} or \textit{operando} experiments may necessitate the refinement of ADPs. These values should be refined in a parametric sense \cite{Stinton-JAC-2007} to constrain their possible values.}, as shown even in this simple mixture created to have minimal overlap.

%%%%
%%%%

\subsection{Peak intensity calibration}

In the use of the \citeasnoun{Hill-JAC-1987b} approach, the scale factor for each phase, as a proxy for peak intensities, is calibrated against the other phases via the unit cell mass and unit cell volume. The unit cell mass can vary through elemental substitution and partial occupancy.  The volume is calculated directly from the unit cell parameters, and their estimates are affected by the calibration of the diffractometer through the zero error, and through the alignment of the specimen through the specimen displacement. Absorption also plays a role, as a less-absorbing sample will have the majority of diffraction away from the specimen surface, resulting in a peak shift akin to specimen displacement.

\subsubsection{Zero error and specimen displacement}

\begin{figure}
\includegraphics{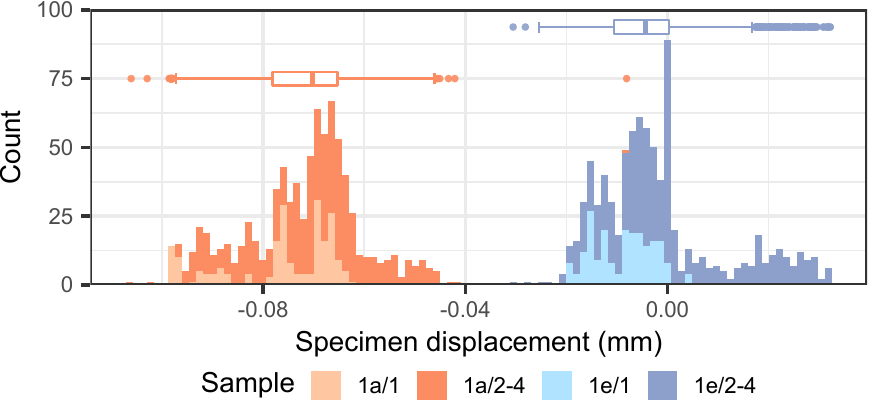}
\caption{Distribution of specimen displacements for samples \numrange{0.01}{0.08}/\numrange{90}{150}. The specimen displacements for refinement type 1 are explicitly shown. }
\label{fig:sd}
\end{figure}

The distributions of specimen displacements for samples 1a and 1e are shown in Figure~\ref{fig:sd} for all intensities, step sizes $\leq$ \SI{0.08}{\degTTh}, and HALs $\geq$ \SI{90}{\degTTh}. There is a clear difference between the two specimens, which were collected at different times in different specimen holders. Interestingly, there is a spread of displacements for refinement type 1, where the unit cell parameters were fixed.

%%%%
%%%%

\begin{figure}
\includegraphics{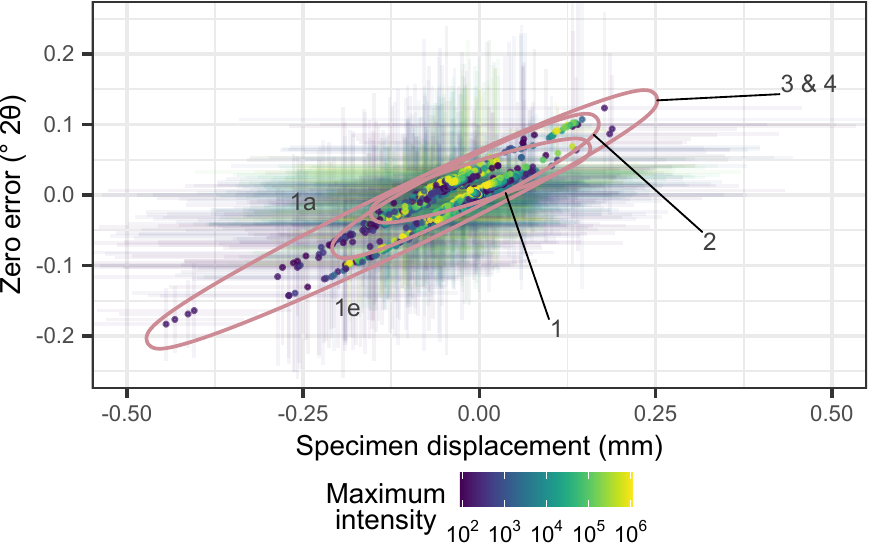}
\caption{Correlation between zero error and specimen displacement for samples \numrange{0.01}{0.08}/\numrange{90}{150} when both parameters are refined concurrently. The range of each refinement type is circled. 1a is the upper line, 1e is the lower line. The error bars represent twice the standard deviation of each estimate, with the standard uncertainty being insignificant on this scale. }
\label{fig:zesd}
\end{figure}

In a separate series of refinements, both specimen displacement and zero error were refined concurrently, showing that there exists an exact correlation between the two parameters -- see Figure~\ref{fig:zesd}. The figure shows two distinct bands associated with samples 1a and 1e, and the different refinement types are circled. There was also a spread of zero error and specimen displacement estimates even for refinement type 1, where the unit cell parameters of all phases are fixed, showing that even this is not enough to stabilise the refinement of both parameters. Taken with the spread of displacement for when the zero error is fixed, this has implications for the use of standard materials in instrument calibration. To this end, zero error and specimen displacement should not be refined together in the model, with the zero error being fixed at a previously determined value, or the diffractometer aligned to remove the error \cite{Cline-JNIST-2015}.

%%%%
%%%%

\subsubsection{Unit cell parameters and volume}

\begin{figure}
\includegraphics{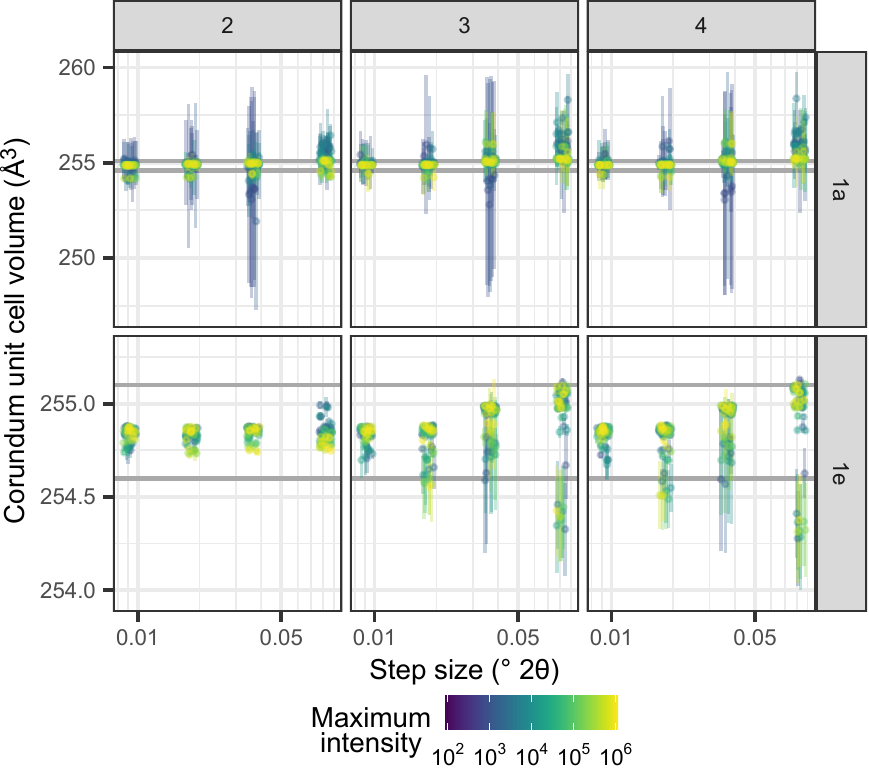}
\caption{Spread of calculated unit cell volume for corundum by refinement type for samples 1a and 1e for all intensities $>$ 500 and all HALs. Error bars represent twice the standard deviation of the estimates. Please note that the vertical axis in the two rows are different; guidelines have been drawn in both rows to indicate an identical vertical range in both rows. The datapoints have been displaced slightly from their x-axis values for clarity.}
\label{fig:vol}
\end{figure}

Individual cell parameters estimates are not of interest, unless they are being used as a proxy for elemental substitution, for example, in Al-substituted goethite \cite{Li-JAC-2012}, rather the unit cell volume is of importance. Figure~\ref{fig:vol} shows the unit cell volume of corundum in samples 1a and 1e for all intensities $>$ 500 and all HALs as a function of step size and refinement type. The error bars displayed on the figure give twice the standard deviation of the volumes from all 200 refinements; the contribution from the standard uncertainty is not significant, showing that the Rietveld esds significantly underestimate the actual spread in estimates.

%%%%
%%%%

\subsection{Weight percentages}

\begin{figure}
\includegraphics{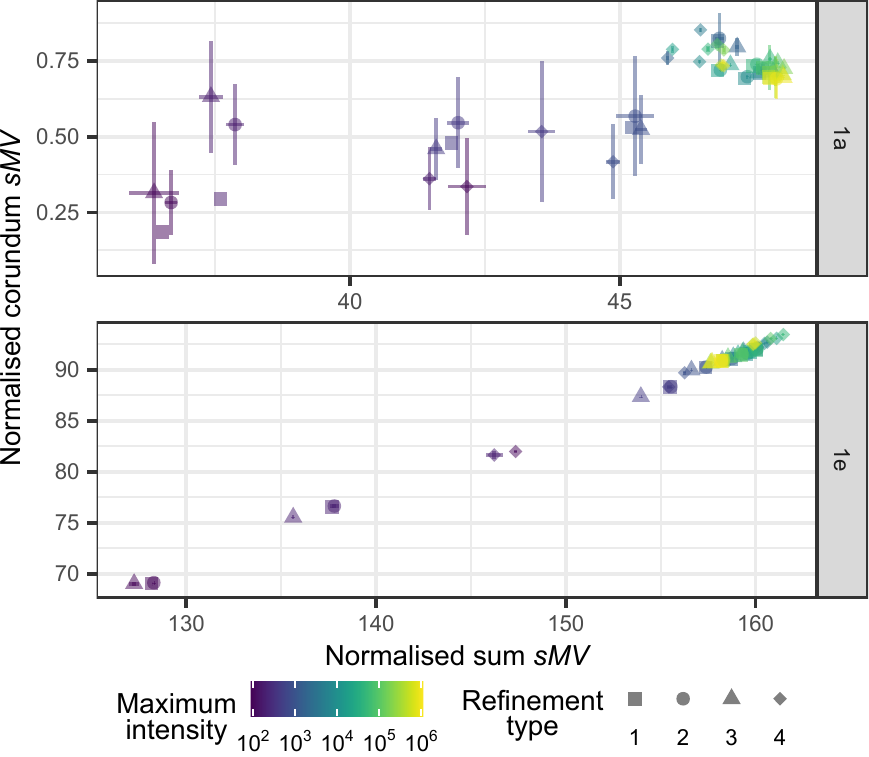}
\caption{Comparison of the $sMV$s for corundum and sum for samples 0.02/110, normalised by maximum intensity. Similar behaviour is exhibited for all refinements. The error bars represent twice the standard deviation.}
\label{fig:smv}
\end{figure}

Weight percentages were calculated by the \citeasnoun{Hill-JAC-1987b} approach through the determination of the product of scale factor, unit cell mass, and unit cell volume -- $sMV$ -- for each phase. Figure~\ref{fig:smv} compares the $sMV$s for corundum and the sum over all phases, and shows that sample 1e has the expected straight-line relationship between $sMV_\mathrm{cor}$ and $sMV_\mathrm{sum}$ through all intensities, whereas no such relationship is present for sample 1a due to the small relative intensities of the corundum peaks.

%%%%
%%%%

\begin{figure}
\includegraphics{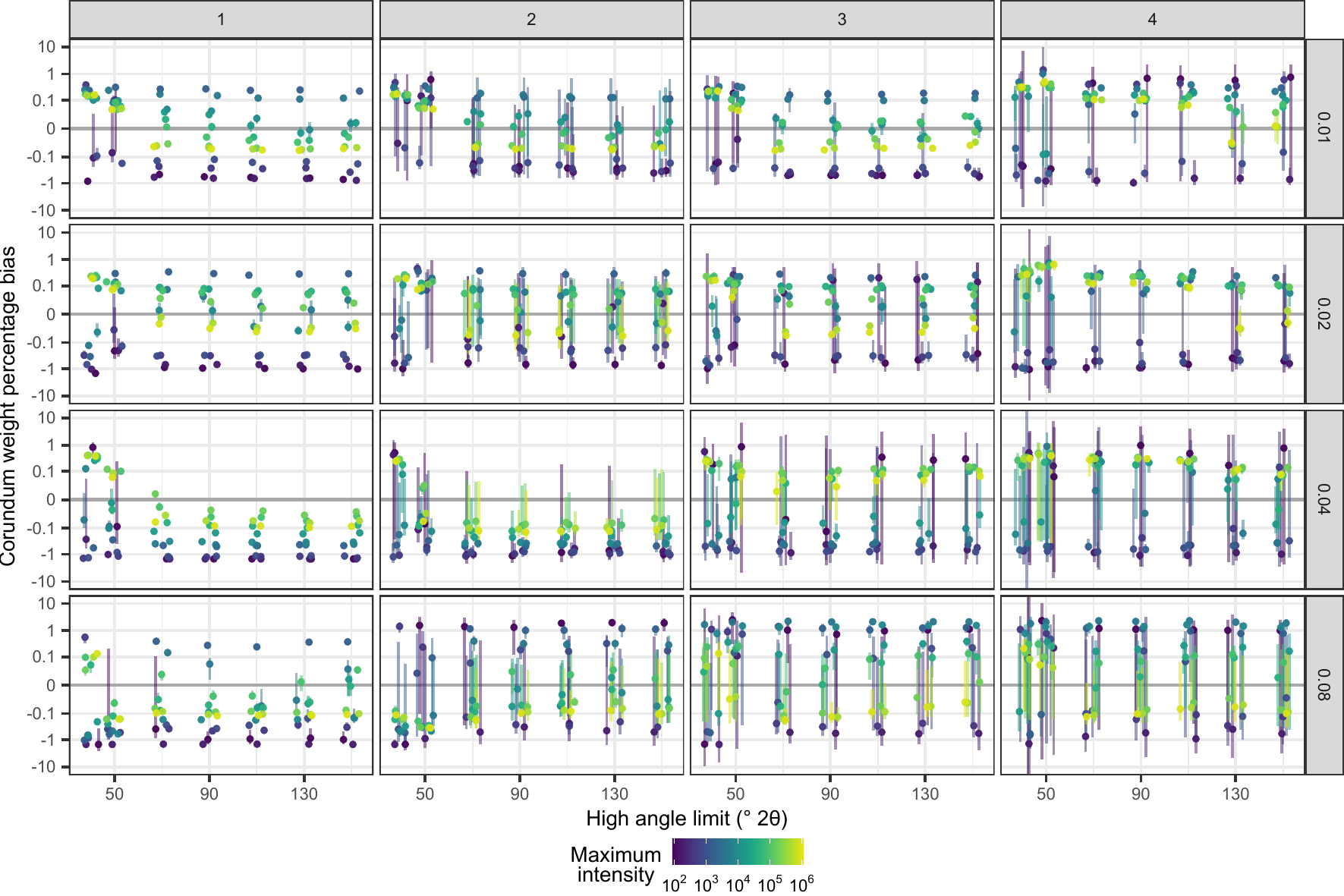}
\caption{Corundum weight percentage bias in sample 1a separated by refinement type and step size. Error bars represent twice the combined standard deviation and uncertainty. Please note that the vertical axis is logarithmic.  For step sizes $>$~\SI{0.08}{\degTTh}, the bias is similar to that of \SI{0.08}{\degTTh}, increasing to $\sim$~\numrange{20}{50} percentage points for \SI{0.32}{\degTTh}. The datapoints have been displaced slightly from their x-axis values for clarity.}
\label{fig:wt1a}
\end{figure}

\begin{figure}
\includegraphics{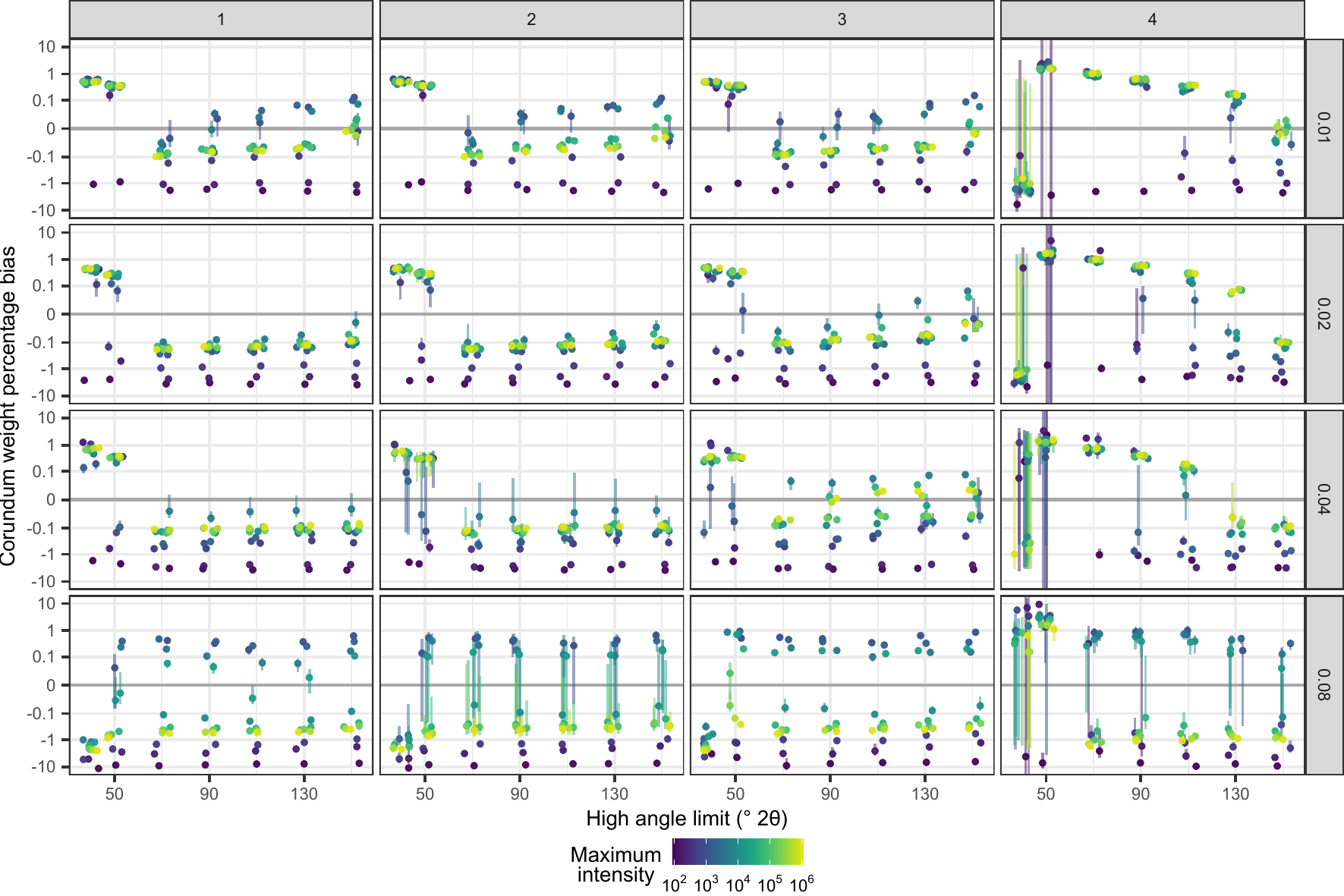}
\caption{Corundum weight percentage bias in sample 1e separated by refinement type and step size. Error bars represent twice the combined standard deviation and uncertainty. Please note that the vertical axis is logarithmic. For step sizes $>$~\SI{0.08}{\degTTh}, the bias is similar to that of \SI{0.08}{\degTTh}, increasing to $\sim 20$ percentage points for \SI{0.32}{\degTTh}. The datapoints have been displaced slightly from their x-axis values for clarity.}
\label{fig:wt1e}
\end{figure}

The weight percentage biases for corundum for all samples, intensities, HALs, refinement types, and step sizes $\le$~\SI{0.08}{\degTTh} are shown in Figures~\ref{fig:wt1a} and \ref{fig:wt1e}. The biases were calculated with respect to the mean of the weight percentages for all maximum intensities $\ge$~\num{20000}, step sizes $=$~\SI{0.01}{\degTTh}, and HAL $=$~\SI{150}{\degTTh}, and therefore shows the effect of the refinement parameters on QPA precision. In both samples, the QPA quickly diverges and increases in error with step sizes beyond \SI{0.08}{\degTTh}, and those step sizes are not longer considered.

%%%%
%%%%

%%%%
%%%%

From an inspection of Figure~\ref{fig:wt1a}, it can be seen that the absolute bias for a great deal of step size $=$~\SI{0.08}{\degTTh} is greater than 1 percentage point; a relative bias of $>$~\SI{65}{\percent}, with quite large errors. Judging by the behaviour of the errors, they are largely due to unit cell parameters, with a large reduction upon moving to refinement type 1. Figure~\ref{fig:wt1e} shows that the QPA from HALs of 40 and \SI{50}{\degTTh} are different from the remaining estimates, reflecting the uneven distribution of diffracted intensity as shown in Figure~\ref{fig:intdist}. This is of particular interest in the quantification of materials with small unit cells, such as metals, as only a few peaks are observed. For all HALs $\geq$~\SI{70}{\degTTh}, the QPA for all intensities is quite flat, with biases remaining approximately constant for refinement types 1 and 2. Refinement type 3 shows a slight decrease in absolute bias with increasing HAL, especially at small step sizes, reflecting the refinement of peakshape parameters. Refinement type 4 shows a large change in bias with HAL, showing the influence of angular range on the refinement of ADPs, which, due to their correlation with the scale factor, affects the QPA \cite{Madsen-ZC-2011}.

The overall performace of each refinement for the entire sample can be summarised using the absolute weighted Kullback-Leibler distance (AwKLD) \cite{Kullback-AMS-1951,Madsen-JAC-2001,Scarlett-JAC-2002}, where
\begin{equation}
\mathrm{wKLD} = 0.01 \times \mathrm{wt\%}_\mathrm{true} \times \ln\qty[\frac{\mathrm{wt\%}_\mathrm{true}}{\mathrm{wt\%}_\mathrm{measured}} ]
\end{equation}

\noindent for each phase, and
\begin{equation}
\mathrm{AwKLD} = \sum_k \abs{\mathrm{wKLD}_k}
\end{equation}
for each refinement; a larger value indicates a ``worse'' refinement in terms of QPA.

\begin{figure}
\includegraphics{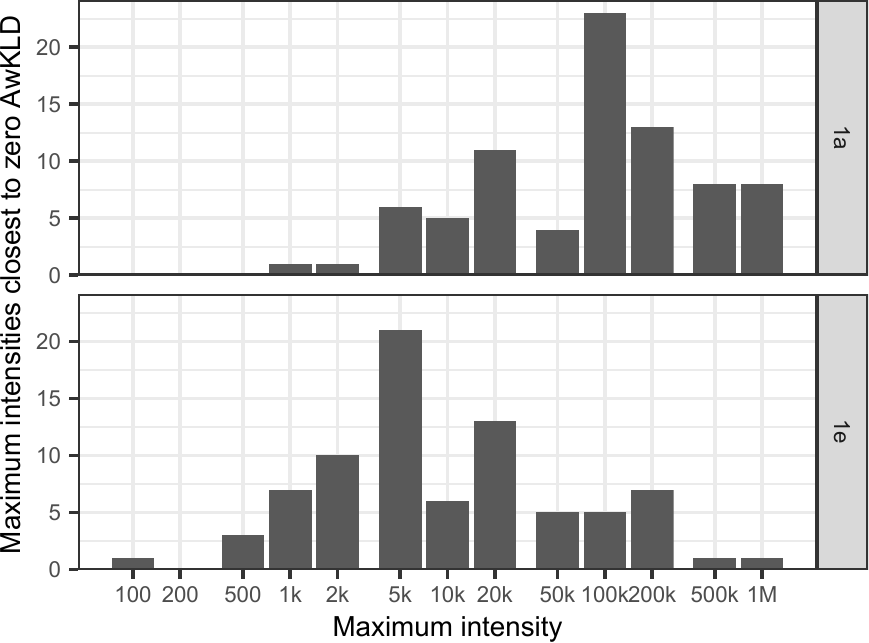}
\caption{Histograms showing the number of diffraction patterns for which that particular maximum intensity had the AwKLD value closest to zero, for samples 1a and 1e. The count is taken over samples $\ge$~70/$\le$~0.08. It can clearly be seen that the sample with phases present in smaller amounts requires maximum intensities much greater than the sample with phases present at approximately equal amounts.}
\label{fig:wKLD_maxint}
\end{figure}

\begin{figure}
\includegraphics{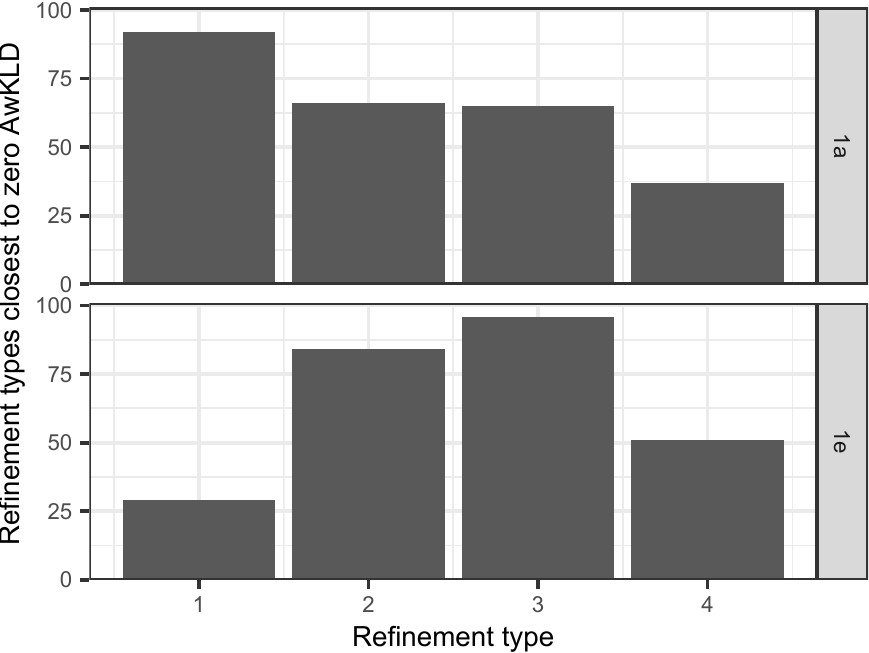}
\caption{Histograms showing the number of diffraction patterns for which that refinement type had the AwKLD value closest to zero, for samples 1a and 1e. The count is taken over samples $\ge$~70/$\le$~0.08. It can clearly be seen that the sample with phases present in smaller amounts requires a more constrained refinement than the sample with phases present at approximately equal amounts, where even this benefits from not refining ADPs.}
\label{fig:wKLD_reftype}
\end{figure}

Two distributions of AwKLD values are shown in Figures~\ref{fig:wKLD_maxint} and \ref {fig:wKLD_reftype}. These figures show the count of maximum intensity and refinement type, respectively, where the corresponding AwKLD value was the closest to zero; i.e. the ``best'' refinement. Figure~\ref{fig:wKLD_maxint} shows that the best maximum intensity is between \numrange{2000}{20000} counts for sample 1e, and between \numrange{20000}{200000} counts for sample 1a, reflecting the nature of each sample, with more counts required to obtain better results for samples containing small weight fractions. Figure~\ref{fig:wKLD_maxint} shows that refinement types 1 and 3 are the best for samples 1a and 1e, respectively. These make sense, as the small peaks present for corundum and zincite in sample 1a could not support the refinement of many parameters, and the systematic change in QPA for refinement type 4 shown in Figure~\ref{fig:wt1e} removes the applicability of ADP refinement in general use.

%%%%
%%%%

%%%%
%%%%

\subsection{Rietveld estimated standard deviations}

The use, calculation, and magnitude of the esds calculated in the Rietveld method are a source of much contention. \citeasnoun{Sakata-JAC-1979} showed that the esds calculated in the least squares process are underestimated due to serial correlation in the residuals, but \citeasnoun{Scott-JAC-1983} showed that the parameter estimates were unbiased.  \citeasnoun{Prince-JAC-1981} and \citeasnoun{Prince-1985} later showed that the esds are correct only if there are no systematic errors in the model or data. Worryingly, \citeasnoun{Scott-JAC-1983} showed that esds can be made arbitrarily small through increasing counting time or decreasing step size. It follows from this that figures-of-merit cannot be used as the sole arbiter of the correctness of a refinement \cite{Toby-PD-2012}, nor esds as measures of the precision of the parameter estimates.

There are numerous proposed approaches to deal with this situation. The simplest is scaling the esd by the GoF \cite{Prince-1995}, although this is not recommended \cite{Schwarzenbach-ACA-1989}. \citeasnoun{Hill-PD-1987} suggested modifying the data collection so that any systematic error is less than the counting statistics, or collecting multiple datasets and using the standard deviation of the now multiple calculated parameter estimates. \citeasnoun{David-JAC-2004} derived an inflation factor to increase structural parameter uncertainties, which depends on the GoF, and the number of data points, model parameters, and peaks.

\begin{figure}
\includegraphics{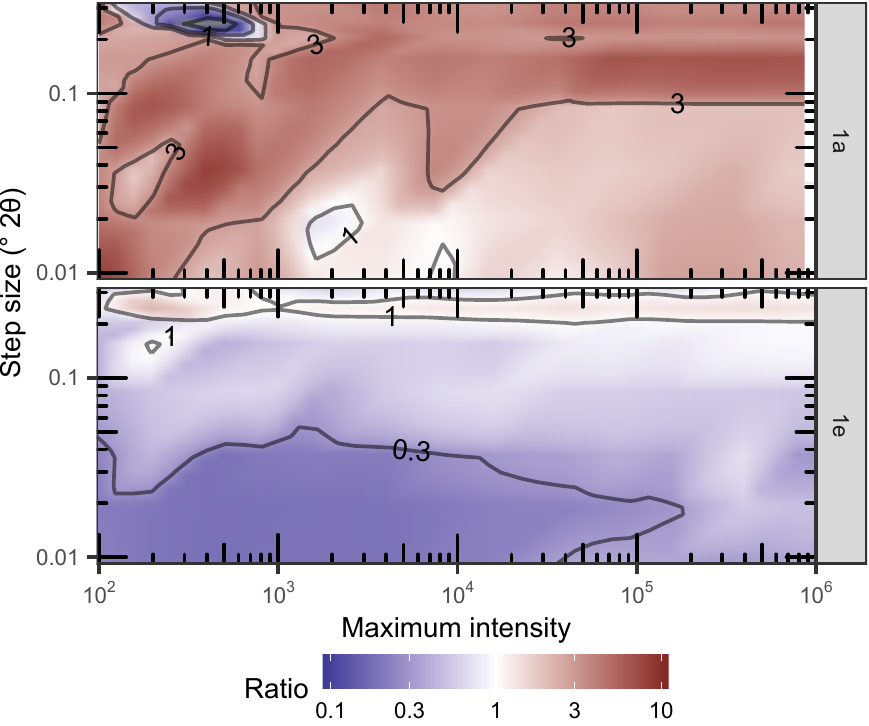}
\caption{The ratio, $\sigma_R/\sigma_T$, of the uncertainties of the weight fraction of corundum from sample 4/130 calculated from this study (Eqn~\ref{eqn:myerror}) and from \citeasnoun{Toraya-JAC-2000} (Eqn~\ref{eqn:torayaerror}), with $D=1$. It is clear that the Toraya method is systematically overestimating errors for evenly distributed intensities, and underestimating for uneven.}
\label{fig:torayaratio}
\end{figure}

\begin{figure}
\includegraphics{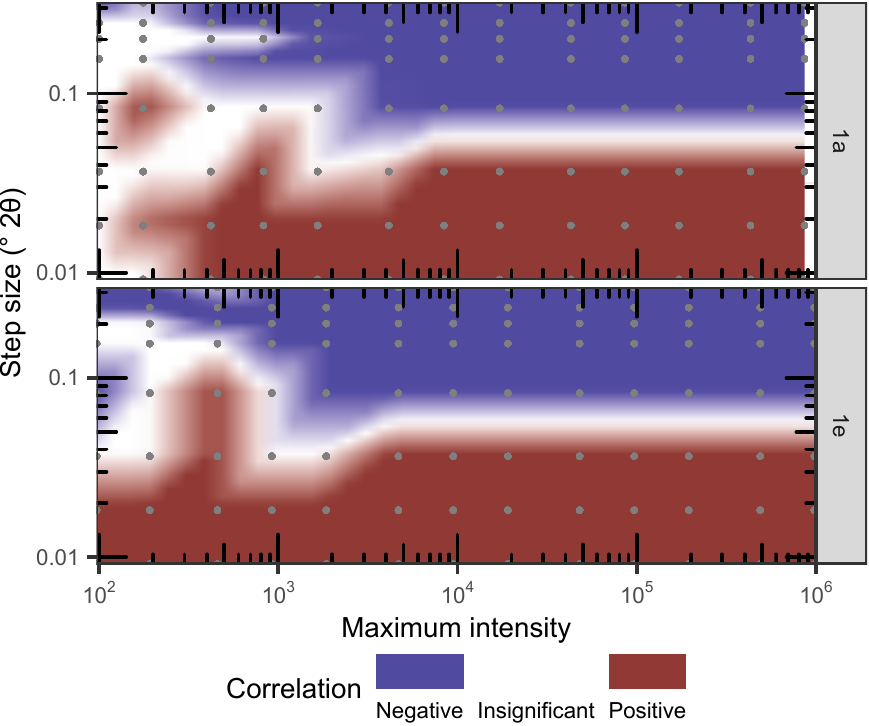}
\caption{Regions of negative, insignificant, and positive serial correlation for sample 4/130 according to Eqns~\ref{eqn:positive} -- \ref{eqn:Q}. The position of the data making the plot is given by the grey points. The overall behaviour of this plot is the same for all others, with large regions of positive and negative serial correlation present at the bottom and top of the plot, the long white bar moving down slightly, and the large white region becoming a little larger.}
\label{fig:goodwDW}
\end{figure}

\citeasnoun{Toraya-JAC-2000} outlined a procedure for estimating the statistical uncertainty of the weight fraction for QPA measurements, with
\begin{equation}
\sigma_T(W_\alpha) = W_\alpha \mathrm{GoF} \sqrt{ \frac{\frac{1}{W_\alpha} - 1}{D \sum y_m^\mathrm{obs}} }
\label{eqn:torayaerror}
\end{equation}
where $0 < D \le 1$ is a measure of peak overlap, with 1 being well-resolved. This uncertainty can be compared to a measure of combined uncertainty from this study, defined by
\begin{equation}
\sigma_R(W_\alpha) = \sqrt{\mathrm{sd}(W_\alpha)^2 + (\mathrm{u}(W_\alpha) \mathrm{GoF})^2   }
\label{eqn:myerror}
\end{equation}
as the ratio $\sigma_R/\sigma_T$, shown in Figure~\ref{fig:torayaratio}. For $D=1$, it is clear that the Toraya method is systematically overestimating errors for evenly distributed intensities and underestimating for uneven intensity distributions by a factor of \numrange{3}{10}. Furthermore, this difference cannot be fully explained by serial correlation, as the variation is inconsistent with the correlations, as shown by the Durbin-Watson statistic in Figure~\ref{fig:goodwDW}.

%%%%
%%%%

The serial correlation of the residuals can be quantitatively assessed using the Durbin-Watson statistic (see Equation~\ref{eqn:d}) \cite{Hill-JAC-1987a,Theil-JASA-1961}, with
\begin{align}
d < Q < 2 & \mathrm{\: \: \: Positive\: serial\: correlation} \label{eqn:positive} \\
d > 4-Q > 2 & \mathrm{\: \: \: Negative\: serial\: correlation} \\
\mathrm{Otherwise}  & \mathrm{\: \: \: Insignificant\: serial\: correlation}
\end{align}
where
\begin{equation}
Q = 2\qty [ \frac{M-1}{M-P} - \frac{3.0902}{\sqrt{M+2}} ]
\label{eqn:Q}
\end{equation}
with the constant representing the \SI{0.1}{\percent} significance point. 

With significant correlation comes over- or underestimation of the esds. The correlations in the models used in this study are summarised by Figure~\ref{fig:goodwDW}. The major features of this figure are the white bar across to high intensity, whose position varies from approximately \SIrange{0.04}{0.06}{\degTTh}, and the large white region at low intensity, whose extent varies from as given, to occupying nearly the entire first third, depending on sample, refinement type, and HAL. It is interesting to note, that while it may be a coincidence, the step size corresponding to insignificant correlation, is approximately the same as the minimum peak FWHM shown in Figure~\ref{fig:fwhm}. Finally, it is striking to see that there are very few refinements where there is no serial correlation; it is not simply a matter of changing the maximum intensity or stepsize to mask the issue, as they would have to be tuned quite carefully. This correlation, and hence mis-estimation of the esds, is not limited to only QPA, and demonstrates that value of an esd rarely has any equivalence to the precision of a parameter estimate.

%%%%
%%%%

Alternative approaches to evaluating the precision of parameter estimates would include
\begin{enumerate}
\item Refining the same data with the same model using different initial parameter values (assesses the impact of model starting values);
\item Collect multiple datasets from the same specimen, emptying and reloading between collections (assesses the impact of data collection and specimen presentation);
\item Use multiple specimens from the same sample (assesses the impact of subsampling and specimen preparation);
\item Collect data using different instruments/geometries (assesses the impact of data collection and models); or
\item Build a more sophisticated analysis model, taking into account more systematic issues such as anisotropic peak broadening, preferred orientation, specimen surface roughness, tube tails, or goniometer eccentricity (reduce serial correlation).
\end{enumerate}
and with combinations of all of the above.

To evaluate the accuracy of QPA, comparison with an independent elemental analysis, such as XRF or ICP-MS, must be carried out; it is not possible to do this solely by diffraction methods.

With all of these failings, esds are useful in determining if a refined parameter is needed in a refinement. If the esd is approximately the same size, or greater than, the estimate to which it is associated, then the validity of that parameter is drawn into question. The user can then apply their domain-specific knowledge to determine whether to either remove the parameter from the model entirely, or to fix it at some physically realistict or meaningful value.

\subsection{Conclusions and recommendations}
  
This study examined the influence of data quality and model type on quantititative phase analysis by the Rietveld method. It found that the data quality, through the maximum intensity, step size, high-angle limit, and the model, through the refinement type, did affect the QPA to varying degrees, as summarised below. Please note that these recommendations have been formed from an ideal mixture, and therefore, are a set of minimum requirements that may need to be changed as specimen complexity increases.

Broadly, for specimens where each phase contributes approximately equally to the diffracted intensity, the pattern's maximum intensity could range between \numrange{1000}{200000} counts above background. The best refinement model was one that did not refine ADPs, but did allow crystallite size/microstrain, unit cell parameters, scale factors, and background to refine. For specimens where there exist minor or trace phases, this intensity range changes to \numrange{5000}{1000000} counts. The best refinement model was one that refined a minimum of parameters. Given that all phases were quite crystalline, step sizes for both types of specimen could range between \SIrange{0.01}{0.04} {\degTTh}, and still yield acceptable results. 

Specifically, from analysis of Figure~\ref{fig:gofrwp}, a maximum intensity of \numrange{20000}{50000} counts will minimise both \Rwp and GoF for the smaller step sizes. Figure~\ref{fig:wKLD_maxint} shows that a maximum intensity of \numrange{2000}{20000} counts for diffraction patterns with an even distribution of intensity between all phases, and up to \numrange{20000}{200000} counts for disparate distributions of intensities is required to minimise the bias in QPA results. \textit{On the whole, data should be taken with a maximum intensity of approximately \num{20000} counts; lower than \num{1000} is not recommended}.

Comparison of Figures~\ref{fig:cryssize} and \ref{fig:pd} with Figure~\ref{fig:fwhm} show that the \textit{step size should be such that there is at least two -- preferably four -- datapoints above the FWHM on the first peak in the pattern}. As peak overlap increases, the step size will need to decrease, along with instrumental parameters, such as equatorial divergence, to increase resolution.

Inspection of Figures~\ref{fig:wt1a} and \ref{fig:wt1e} in the context of Figure~\ref{fig:intdist} shows that there is minimal bias in the quantification after a high-angle limit of \SI{70}{\degTTh} due to a consistent increase in cumulative intensity with angle, showing that \textit{data should be collected from the first peak to a point where there is a constant increase in cumulative intensity with angle, to a point where peaks no longer appear, or to the highest practicable angle possible on the goniometer}. This last point is of particular importance if atomic displacement parameters are to be refined. Care should be taken that the incident beam is contained entirely within the specimen, and doesn't overspill onto the holder or other parts of the instrument.

In terms of the refinement model, \textit{zero error and specimen displacement should not be refined together in the same refinement}. Their correlation is such that even with fixed unit cell parameters, they are still able to refine to different values -- see Figure~\ref{fig:zesd} -- which impacts on the use of standard materials for instrument calibration. Specifically, the zero error should be fixed at a known value, or the diffractometer aligned to remove it. \textit{Atomic displacement parameters should not be refined at all for minor phases, and not for overlapping phases} -- see Figure~\ref{fig:tp}. The relevant literature and databases should be consulted for relevant values derived from single-crystal measurements. If these parameters must be refined, then diffraction data should be collected to the highest angle possible with an appropriately small step size, as indicated by the variation in parameter estimates in Figure~\ref{fig:tp2} and by the systematic change in QPA in refinement type 4 with high-angle limit in Figure~\ref{fig:wt1e}. Finally, as many parameters as possible should be fixed for minor phases, or their range significantly limited by fixed limits, as there won't be sufficient data quality to support their refinement. If these minor phases are critical, then they should be concentrated and analysed to allow their parameters to be determined and then applied to the original sample.

It is hoped that this study, and its recommendations, will aid new QPA-by-Rietveld users to collect and analyse data to a high standard, and provide a foundation for on-going learning in the application of the Rietveld method and diffraction analysis in general.

     %-------------------------------------------------------------------------
     % The back matter of the paper - acknowledgements and references
     %-------------------------------------------------------------------------

     % Acknowledgements come after the appendices

\appendix

\section{Determination of parameter refinement order}

To evaluate the effect of parameter refinement order on model outcomes, 26 different models were constructed with parameters being introduced and refined in different manners. All initial parameter values were set in the same manner for all models – see Table~\ref{table:prmorderstart}. Parameters were fixed, refined, or introduced to the model in accordance with Table~\ref{table:prmorder}. In general, parameter refinement order was chosen under the general provision ``making the biggest difference first'' \cite{Toby-VolH-2019}, by obtaining correct peak intensities, peak positions, and finally, correct peak shapes. Atomic displacement parameters were refined last, as they depend on subtleties of peak intensity. All models were refined using TOPAS Academic v6 \cite{Coelho-JAC-2018a}. 

The data used in the determination of the parameter refinement order was that provided by the IUCr Commission on Powder Diffraction round robin on quantitative phase analysis \cite{Madsen-JAC-2001} for sample 1e. The data collection and instrumental details are available therein. 

\begin{figure}
\includegraphics{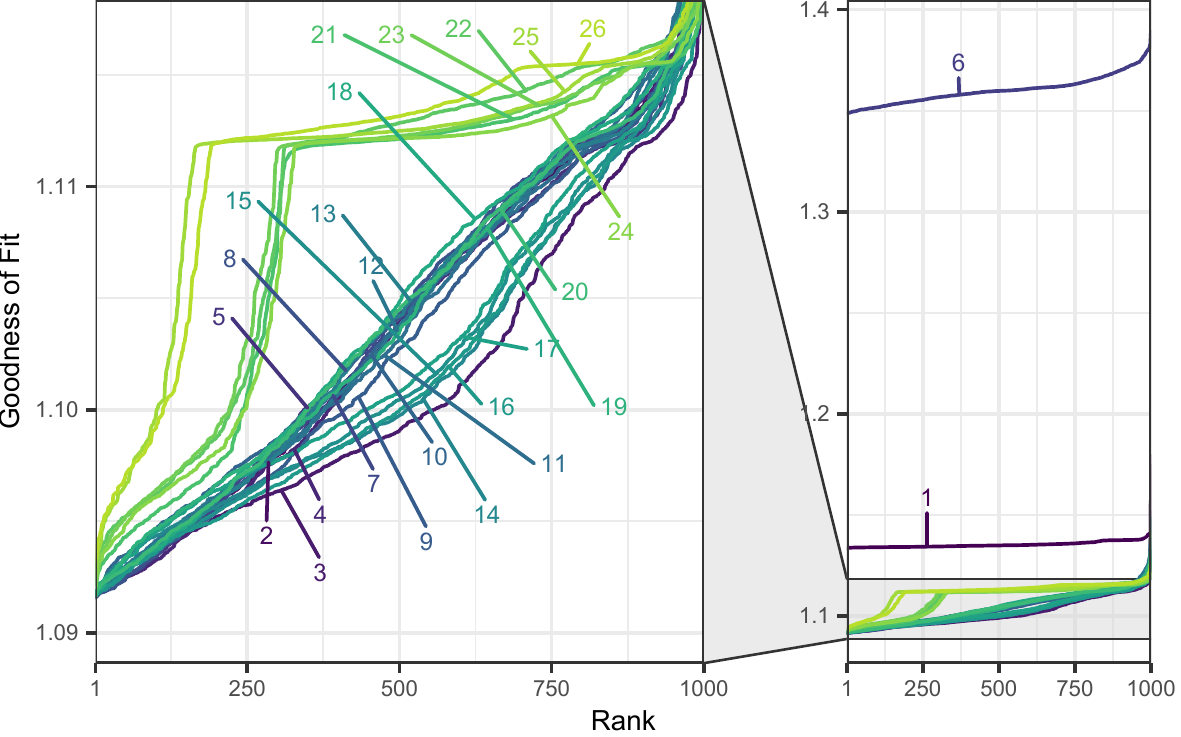}
\caption{Goodness of Fit values of the various parameter order models in rank order.}
\label{fig:prmorder}
\end{figure}

1000 complete refinements of each model were carried out, and the parameter estimates, derived parameter estimates, estimated standard deviations, and figures-of-merit, were written to file upon completion of the final step of each complete refinement. The GoFs of the models are plotted in rank order in Figure~\ref{fig:prmorder}. The models fall into five groups, from top to bottom, then left to right in the figure: (i) 6, (ii) 1, (iii) 26,25,24,23,22,21, (iv) 20,18,7, 4,11,8,19,5,13,2,10,9,12, and (v) 14,15,17,16,3. 

%%%%%%%%%%%%
\begin{table}
\caption{Model starting parameter values and refinement limits. Corundum, fluorite, and zincite are abbreviated as cor, flu, and zin. Numbers in brackets indicate a number chosen at random, uniformly the two given values.}
\label{table:prmorderstart}
\begin{center}
\begin{tabular}{rrrcc}
\multicolumn{3}{c}{Parameter}                                   & Symbol                & Value/range      \\
\midrule
\multicolumn{3}{r}{Chebyshev background}                                          & bkg                   & 0 0 0 0 0        \\
\multicolumn{3}{r}{1/X background}                                         & bkg                   & 10000            \\
\multicolumn{3}{r}{Zero error (\si{\degTTh})}                                        & ZE                    & 0                \\
\multicolumn{3}{r}{Specimen displacement (\si{\milli\m})}                                        & SD                    & 0                \\
\multicolumn{3}{r}{Packing density}                                     & PD                    & (0.2, 0.7)       \\
\multirow{3}{*}{Scale factor}        & \multicolumn{2}{r}{cor}  & \multirow{3}{*}{sc}   & (0.00007, 0.007) \\
                                     & \multicolumn{2}{r}{flu}  &                       & (0.0011, 0.011)  \\
                                     & \multicolumn{2}{r}{zin}  &                       & (0.0038, 0.038)  \\
\multirow{5}{*}{Unit cell parameter (\si{\angstrom})} & \multirow{2}{*}{cor} & a & \multirow{5}{*}{cell} & (4.75, 4.77)                \\
                                     &                      & c &                       & (12.94, 13.04)                \\
                                     & flu                  & a &                       & (5.44, 5.48)                \\
                                     & \multirow{2}{*}{zin} & a &                       & (3.24, 3.26)                \\
                                     &                      & c &                       & (5.19, 5.23)                \\
\multicolumn{3}{r}{Lorentzian crystallite size (\si{\nano\m})}                                         & csL                   & (50, 500)        \\
\multicolumn{3}{r}{Gaussian crystallite size (\si{\nano\m})}                                         & csG                   & (50, 500)        \\
\multicolumn{3}{r}{Lorentzian microstrain}                                         & strL                  & (0.001, 0.5)     \\
\multicolumn{3}{r}{Gaussian microstrain}                                         & strG                  & (0.001, 0.5)     \\
\multicolumn{3}{r}{Isotropic atomic displacement parameter (\si{\angstrom\squared})}                                      & B                     & (0.3,1)         
\end{tabular}
\end{center}
\end{table}
%%%%%%%%%%%%%%%%

%%%%%%%%%%%%
\begin{table}
\caption{Parameter refinement models showing the order in which various parameters were added and refined in the quantitative phase analysis model. In all cases, parameters in previous steps were co-refined with parameters in the current step.}
\label{table:prmorder}
\begin{center}
\resizebox{\textwidth}{!}
{\begin{tabular}{ccccccccccc}
\multirow{2}{*}{\makecell{Model\\number}} & \multicolumn{10}{c}{Refinement step}   \\
                              & 1 & 2 & 3 & 4 & 5 & 6 & 7 & 8 & 9 & 10 \\
\midrule
1  & All               & All                    &                   &             &          &          &          &          &     &    \\
2  & bkg sc            & cell                   & ZE                & SD          & csL      & PD       & strG     & csG strL & B   & All \\
3  & bkg sc            & cell                   & ZE SD             & csL         & PD       & strG     & csG strL & B        & All	&      \\
4$^1$  & bkg sc            & cell                   & ZE                & csL         & PD       & strG     & csG strL & B        & All	&    \\ 
5$^2$   & bkg sc            & cell                   & SD                & csL         & PD       & strG     & csG strL & B        & All	&     \\	 
6$^{1,2}$    & bkg sc            & cell                   & csL               & PD          & strG     & csG strL & B        & All      &     &	 	 \\
7$^1$   & bkg sc            & cell ZE                & csL               & PD          & strG     & csG strL & B        & All      &     &	 	 \\	 	 
8$^2$   & bkg sc            & cell SD                & csL               & PD          & strG     & csG strL & B        & All      &     &	 	 \\	 	 
9  & bkg sc            & cell ZE SD             & csL               & PD          & strG     & csG strL & B        & All      &     &	 	 \\	 	 
10 & bkg sc            & cell ZE SD             & csL strG          & PD          & csG strL & B        & All      &          &     &	 	 \\	 	 	 
11 & bkg sc            & cell ZE SD             & cs str            & PD          & B        & All      &          &          &     &	 	 \\	 	 	 	 
12 & bkg sc            & cell ZE SD             & cs str PD         & B           & All      &          &          &          &     &	 	 \\		 	 	 	 	 
13 & bkg sc            & cell                   & SD                & ZE          & csL      & PD       & strG     & csG strL & B   & All \\
14 & bkg sc            & cell                   & ZE SD             & csL strG    & PD       & csG strL & B        & All      &     &    \\		 	 
15 & bkg sc            & cell                   & ZE SD             & cs str      & PD       & B        & All      &          &     &    \\	 
16 & bkg sc            & cell                   & ZE SD             & cs str PD   & B        & All	     &          &          &     &    \\ 	 	 	 
17 & bkg sc            & cell                   & ZE SD             & cs str PD B & All      &    	     &          &          &     &    \\ 	 	 	 	 	 
18 & bkg sc            & cell                   & ZE SD cs str PD B & All	 	 	 &          &    	     &          &          &     &    \\ 		 	 	 
19 & bkg sc            & cell ZE SD             & cs str PD B       & All	 	 	 &          &    	     &          &          &     &    \\	 	 	 	 	 	 
20 & bkg sc            & cell ZE SD cs str PD B & All	 	 	        &    	 	 	 &          &    	     &          &          &     &    \\	      	 	 	 	 
21 & bkg sc cell       & ZE SD                  & csL               & PD          & strG     & csG strL & B        & All      &     &    \\		 	 
22 & bkg sc cell       & ZE SD                  & csL strG          & PD          & csG strL & B        & All      &          &     &    \\		 	 	 
23 & bkg sc cell       & ZE SD                  & cs str            & PD          & B        & All      &          &          &     &    \\	 	 	 	 
24 & bkg sc cell       & ZE SD                  & cs str PD         & B           & All      &          &          &          &     &    \\		 	 	 	 	 
25 & bkg sc cell ZE SD & cs str PD              & B                 & All         &          &          &          &          &     &    \\	 	 	 	 	 	 
26 & bkg sc cell ZE SD & cs str PD B            & All               &             &          &          &          &          &     &    \\	 	
\end{tabular}}

\end{center}
\footnotesize{$^1$ Specimen displacement fixed at zero. $^2$ Zero error fixed at zero.}
\end{table}
%%%%%%%%%%%%%%%%

%%%%
%%%%%%

From inspection of this figure, it is apparent that the best GoF is 1.091. It is interesting to note that Model 1, were all parameters were refined simultaneously does not produce the best refinement. Model 6, with specimen displacement and zero error fixed at zero is extremely bad, showing that at least one of these values is incorrect. All other models are capable of reaching this ‘best’ value, with varying degrees of probability. All groups have a few complete refinements with quite relatively high GoFs, which then come down to a plateau (except for groups iv and v), followed by a continuous decrease to the lowest GoF.

The main distinguishing points in models between these five groups are that in group i, zero error and specimen displacement are fixed at zero; group ii, all parameters are refined together; group iii, background coefficients and scale factors are refined together with cell parameters in the first step; and in group v, zero error and specimen displacement are refined together in the same refinement step by themselves. 

The model with the consistently lowest GoF is 3, and this was chosen as the basis for the parameter refinement order to be used for the robustness study.

\ack{I would like to acknowledge helpful discussions and inspiration from Ian Madsen and Nikki Scarlett (CSIRO). IM also gave critcial feedback on the final manuscript. Diffraction data were collected using the X-ray instrumentation (ARC LE0775551) at the John de Laeter Centre, Curtin University.}

     % References are at the end of the document, between \begin{references}
     % and \end{references} tags. Each reference is in a \reference entry.
\referencelist[row]

\end{document}

%\preprint{}

\title{Supplementary information -- The effect of data quality and model parameters on the quantitative phase analysis of X-ray diffraction data by the Rietveld method}


\author{Matthew R. Rowles}
\email[]{matthew.rowles@curtin.edu.au}
%\thanks{}
\affiliation{John de Laeter Centre, Curtin University, Perth, WA Australia}


%\date{\today}

%\begin{abstract}
%This supplementary information presents the results obtained from the models including a refining zero error in conjunction with a specimen displacement. All other information about the models, data, refinement methods, etc, is as given in the main paper.
%\end{abstract}

%\keywords{}

\maketitle
\section{}
\section{Supplementary tables and figures}

This section gives supplementary tables and figures to the main paper.

%%%%%%%%%%%%%%%%
\begin{table}[h]
\caption{Data collection conditions for the diffraction patterns of varying intensity.}
\label{table:collection}
\begin{center}
\begin{tabular}{S[table-format=6.0] r c S[table-format=1.6] S[table-format=1.6] c c c c}      % Alignment for each cell: l=left, c=center, r=right
\toprule
{\makecell{Nominal \\ maximum \\ intensity}} & 
\multicolumn{2}{c}{\makecell{Tube \\ current \\ (\si{\milli\ampere})}} & 
\multicolumn{2}{c}{\makecell{Step size \\ (\si{\degTTh})}} & 
\multicolumn{2}{c}{\makecell{Count \\ time per \\ step (\si{\second})}} & 
\multicolumn{2}{c}{\makecell{Datasets \\ collected}} \\
             & 1a          & 1e         & {1a}          & {1e}         & 1a          & 1e         & 1a          & 1e         \\
\midrule
100   	& 5  	&     	& 0.004585	&                 	& 0.01	&     	& 1  	&    \\
100   	& 6  	& 10	& 0.004585	& 0.004578	& 0.01	& 0.01	& 1  	& 2  \\
300   	& 8  	& 10	& 0.009171	& 0.009155	& 0.02	& 0.03	& 2  	& 1  \\
500   	& 14	& 10	& 0.009171	& 0.009155	& 0.02	& 0.05	& 2  	& 1  \\
1000  	& 11	& 10	& 0.009171	& 0.009155	& 0.05	& 0.10	& 1  	& 1  \\
3000  	& 11	& 20	& 0.009171	& 0.009155	& 0.15	& 0.15	& 1  	& 1  \\
5000  	& 12	& 20	& 0.009171	& 0.009155	& 0.23	& 0.25	& 1  	& 1  \\
10000 	& 24	& 40	& 0.009171	& 0.009155	& 0.23	& 0.25	& 1  	& 1  \\
30000 	& 36	& 40	& 0.009171	& 0.009155	& 0.46	& 0.75	& 1  	& 1  \\
50000 	& 40	& 40	& 0.009171	& 0.009155	& 0.69	& 1.25	& 1  	& 1  \\
100000	& 40	& 40	& 0.009171	& 0.009155	& 1.38	& 2.50	& 9  	& 9  \\
\bottomrule
\end{tabular}
\end{center}
\end{table}
%%%%%%%%%%%%%%%%

%%%%
\begin{figure}[h]
\includegraphics{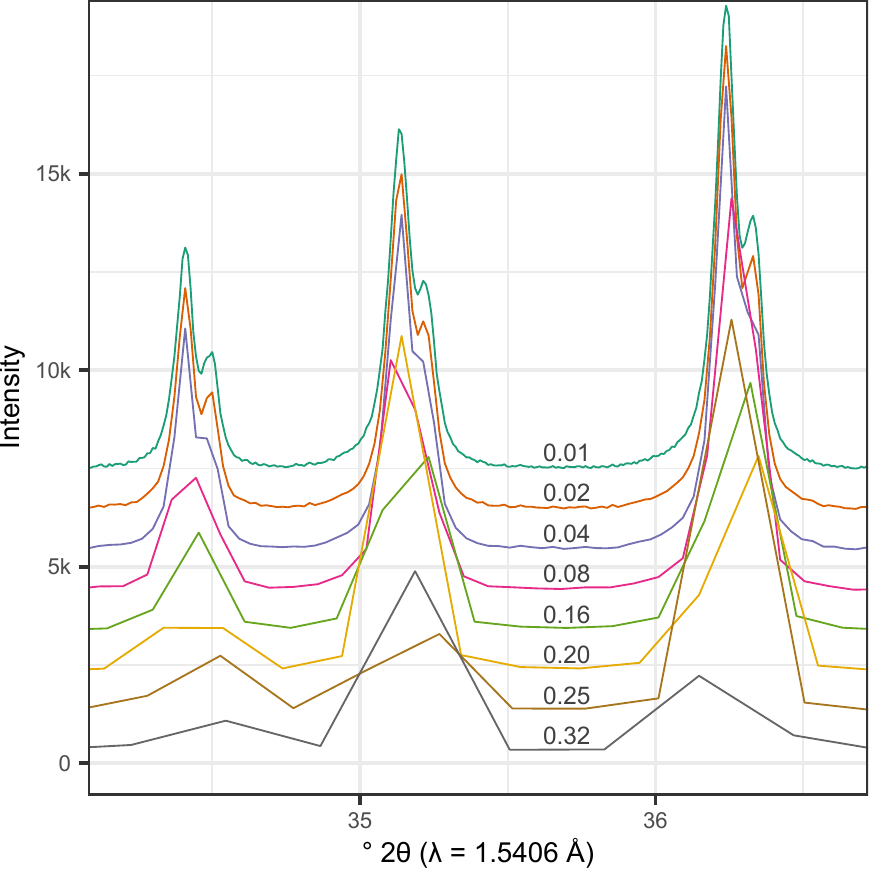}
\caption{Diffraction data of sample 1e for all step sizes with a nominal maximum intensity of \num{20000} counts. To give diffraction patterns with different step sizes, points were dropped from the original measured data. The patterns are displaced vertically for clarity.}
\label{fig:stepsize}
\end{figure}

%%%%

%%%%%%%%%%%%%%%%
\begin{table}[h]
\caption{Nominal and actual (a) step sizes (\si{\degTTh}) and (b) maximum intensities (counts) of the diffraction patterns used in the modelling. Every $n^\mathrm{th}$ datapoint was kept to construct each of the diffraction patterns.}
\label{table:nomact}
\begin{center}
(a)

\begin{tabular}{S[table-format=2.0] S[table-format=1.2] S[table-format=1.6] S[table-format=1.6] }  
\toprule
n	& Nominal	& {1a}       	& {1e}    \\
\midrule
1	& 0.01   	&0.009171	& 0.009155 \\
2	& 0.02   	&0.01834	& 0.01831   \\
4	& 0.04   	&0.03668	& 0.03662   \\
9	& 0.08   	&0.08254	& 0.08239   \\
17	& 0.16   	&0.1559 	& 0.1556     \\
22	& 0.20   	&0.2018 	& 0.2014     \\
27	& 0.25   	&0.2476 	& 0.2472     \\
35	& 0.32   	&0.3210  	& 0.3204     \\
\bottomrule
\end{tabular}

\bigskip

(b)

\begin{tabular}{S[table-format=7.0] S[table-format=6.0] S[table-format=6.0] | S[table-format=7.0] S[table-format=6.0] S[table-format=6.0]  }
\toprule
{Nominal}	& {1a}  	& {1e}  &   {Nominal}	& {1a}  	& {1e} \\
\midrule
100   	& 100   	& 97     & 20000  	& 17200 	& 19100  \\
200   	& 176   	& 192    & 50000 	& 42900 	& 48000  \\
500   	& 421   	& 458   & 100000	& 85900 	& 96700  \\
1000  	& 826   	& 921   & 200000	& 172000	& 195000 \\
2000  	& 1660  	& 1850  & 500000	& 431000	& 488000 \\
5000  	& 4140  	& 4680  & 1000000	& 862000	& 978000 \\
10000 	& 8400  	& 9460  &                  &                &             \\
\bottomrule
\end{tabular}
\end{center}
\end{table}
%%%%%%%%%%%%%%%%

\section{Refining a zero error}

Here are figures and tables derived from the models with a refining zero error in conjunction with a specimen displacement. All other information about the models, data, refinement methods, etc, is as given in the main paper.

%%%%%%%%%%%%
\begin{table}
\caption{Parameter values used when, according to the refinement type, a given parameter was fixed.}
\label{table:SI_modelfinish}
\begin{center}
\begin{tabular}{rrrS[table-format=3.6]}
\toprule
\multicolumn{3}{c}{Parameter}                                    & {Value}    \\
\midrule
\multicolumn{3}{r}{Packing density}                              & 0.172    \\
\midrule
\multirow{3}{*}{\makecell{Crystallite size \\ Lorentzian (\si{\nano\m})}}    & \multicolumn{2}{r}{Cor}   & 313 \\
                                     & \multicolumn{2}{r}{Flu}   & 590      \\
                                     & \multicolumn{2}{r}{Zin}   & 293      \\
\midrule
\multirow{2}{*}{\makecell{Microstrain \\ Lorentzian}}         & \multicolumn{2}{r}{Cor}   & 0.0195   \\
                                     & \multicolumn{2}{r}{Flu}   & 0.0415   \\
\midrule
\multirow{5}{*}{\makecell{Unit cell \\ parameter (\si{\angstrom})}} & \multirow{2}{*}{Cor} & a  & 4.75953 \\
                                     &                      & c  & 12.99328 \\
                                     & Flu                  & a  & 5.46463 \\
                                     & \multirow{2}{*}{Zin} & a  & 3.25005 \\
                                     &                      & c  & 5.20697  \\
\midrule
\multirow{6}{*}{\makecell{Isotropic atomic \\ displacement \\ parameter (\si{\angstrom\squared})}}   & \multirow{2}{*}{Cor} & Al & 0.249    \\
                                     &                      & O  & 0.193    \\
                                     & \multirow{2}{*}{Flu} & Ca & 0.468    \\
                                     &                      & F  & 0.692    \\
                                     & \multirow{2}{*}{Zin} & Zn & 0.524    \\
                                     &                      & O  & 0.321   \\
\bottomrule
\end{tabular}
\end{center}
\end{table}
%%%%%%%%%%%%

%%%%
\begin{figure}
\includegraphics{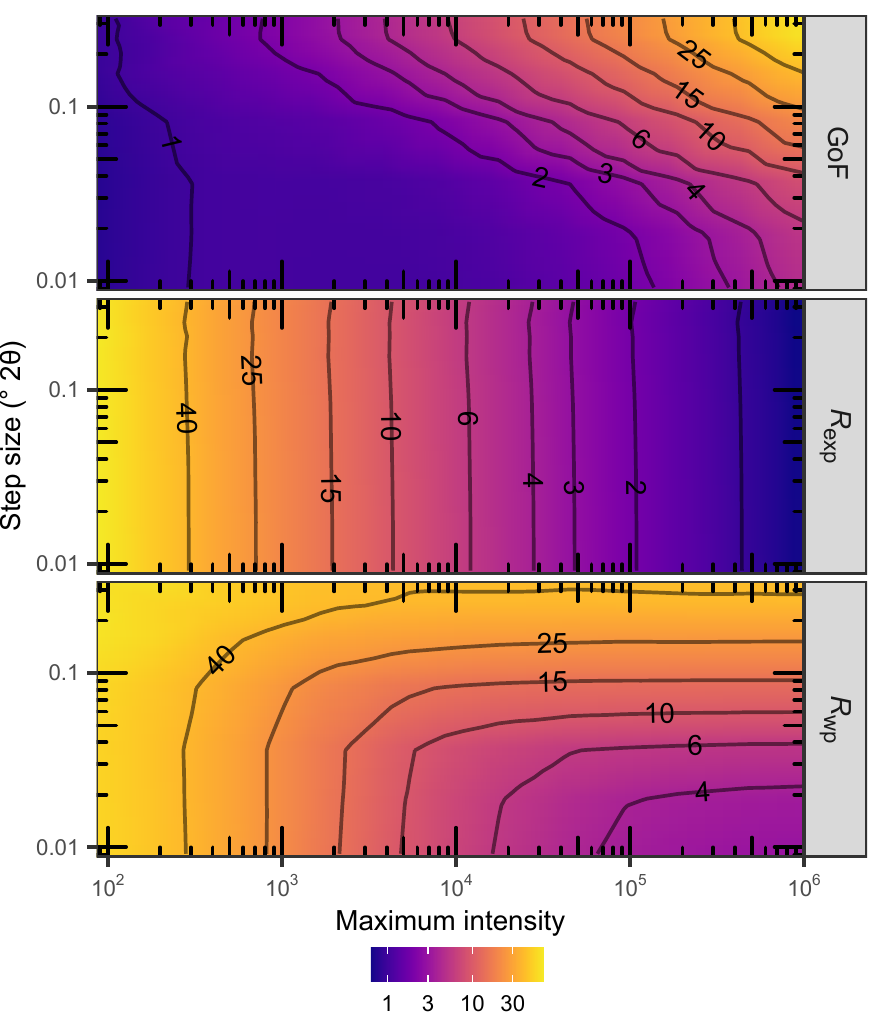}
\caption{The figures-of-merit, GoF, \Rexp, and \Rwp, for sample 1e/4/150. The trends evident in these plots are repeated throughout all the models. It can be seen that the desire for a low Gof and \Rwp are at odds with each other with respect to maximum intensity.}
\label{fig:SI_fom}
\end{figure}
%%%%

%%%%
\begin{figure}
\includegraphics{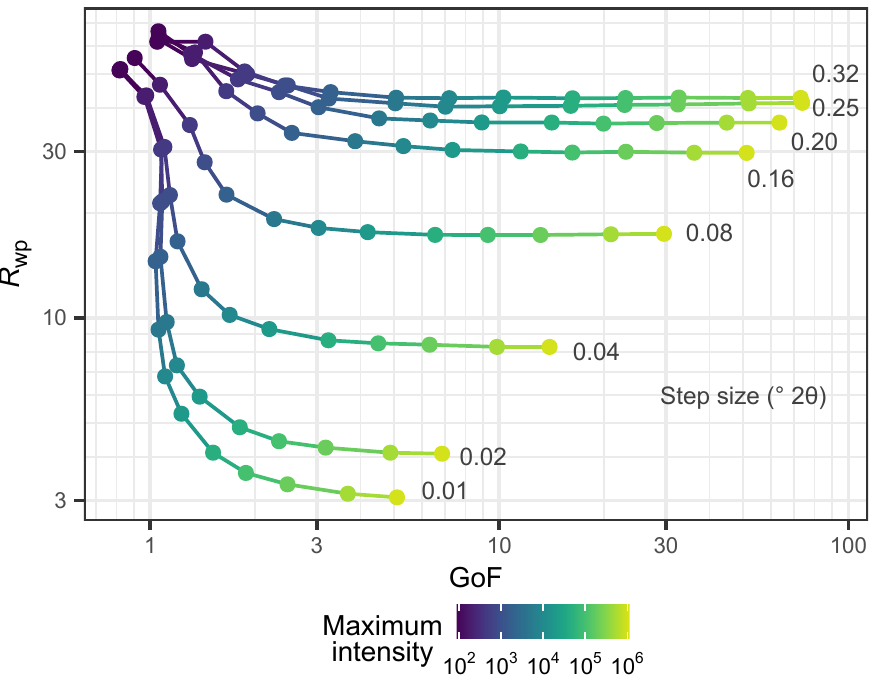}
\caption{Comparison of \Rwp and GoF for all sample 1e averaged over all refinement types and HALs, showing that step size is the best predictor low \Rwp and GoF, when coupled with a maximum intensity of $\sim$ \numrange{20}{50000} counts.}
\label{fig:SI_gofrwp}
\end{figure}
%%%%

%%%%%
\begin{figure}
\includegraphics{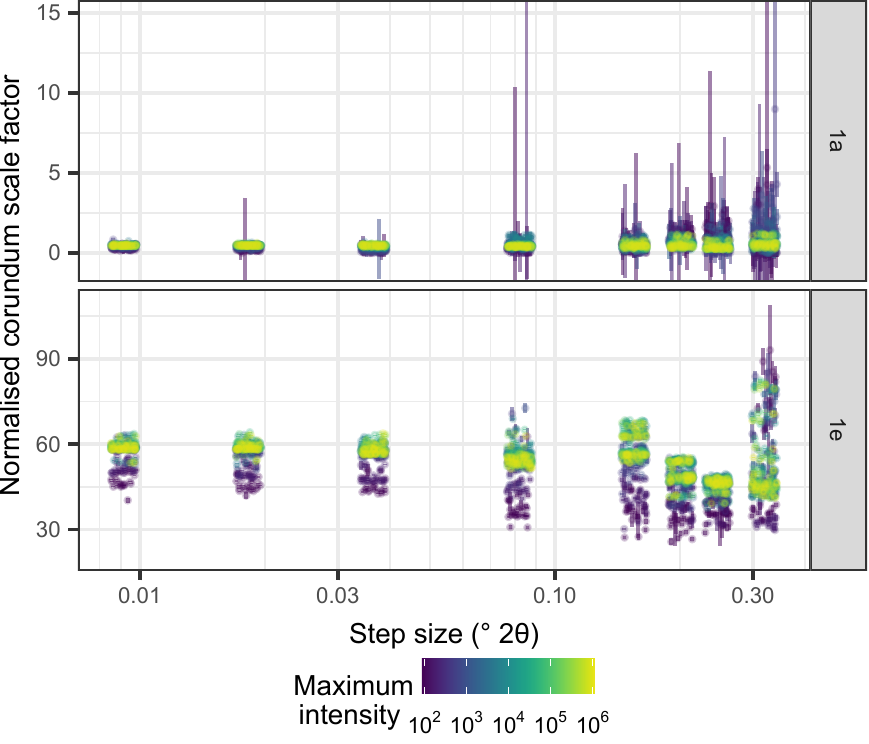}
\caption{Scale factors for corundum normalised by maximum intensity for samples 1a and 1e. The error bars represent twice the standard uncertainty. The datapoints have been displaced slightly from their x-axis values for clarity.}
\label{fig:SI_sf}
\end{figure}
%%%%

%%%%
\begin{figure}
\includegraphics{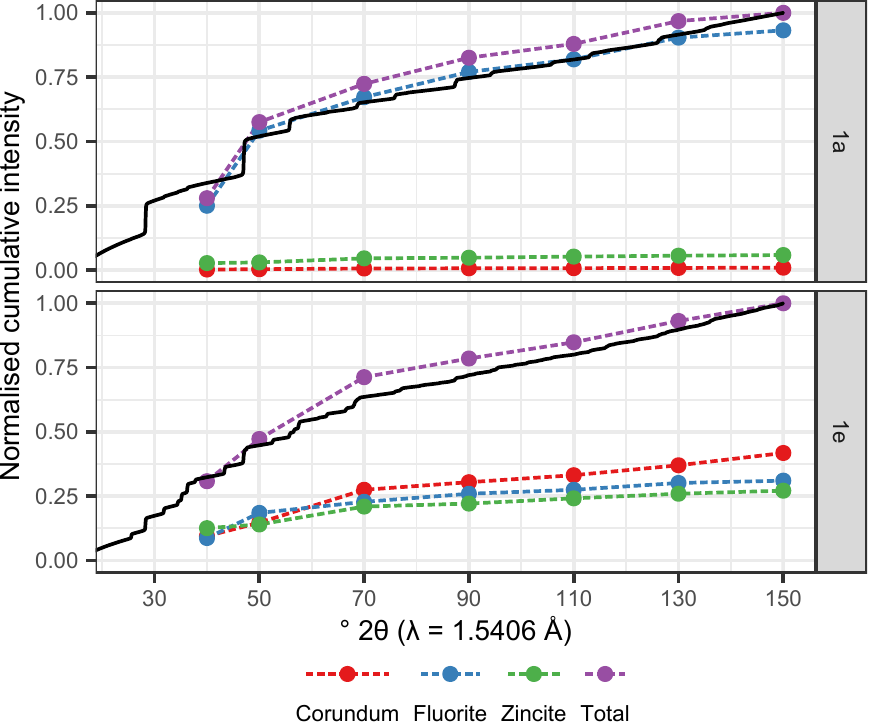}
\caption{The cumulative intensity of the diffraction data presented in Figure~\ref{fig:diffdata} and the numerical area of the individual phases and their sum normalised to the total diffracted intensity. In both samples 1a and 1e, it can be seen that intensities after \SI{70}{\degTTh} are evenly distibuted. The areas attributed to each phase change in relative distribution with low HAL values for both samples; after the intensities stabilise at \SI{70}{\degTTh}, their relative contributions remain constant.}
\label{fig:SI_intdist}
\end{figure}
%%%%

%%%%
\begin{figure}
\includegraphics{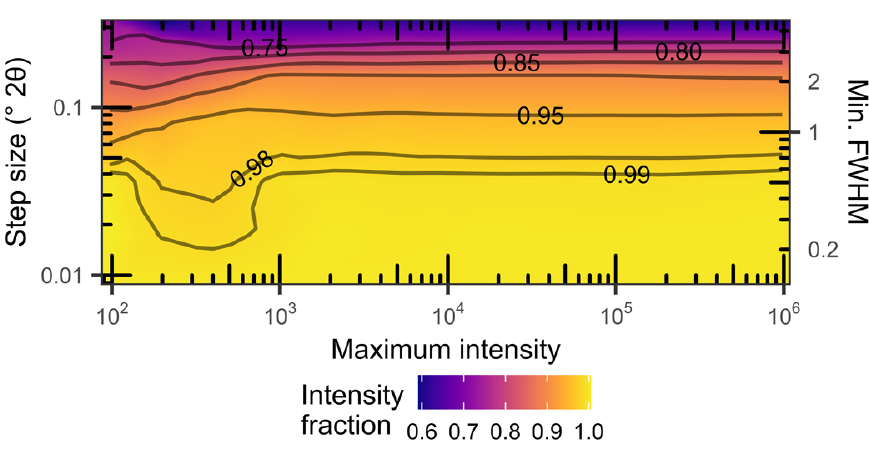}
\caption{Fraction of total intensity in sample 1e/3/150 as a ratio of the total intensity present in the \SI{0.01}{\degTTh} step size pattern. The right axis gives the step size as a function of the average FWHM of the first peak, as given in Figure~\ref{fig:fwhm}.}
\label{fig:SI_numarea}
\end{figure}
%%%%

%%%%
\begin{figure}
\includegraphics{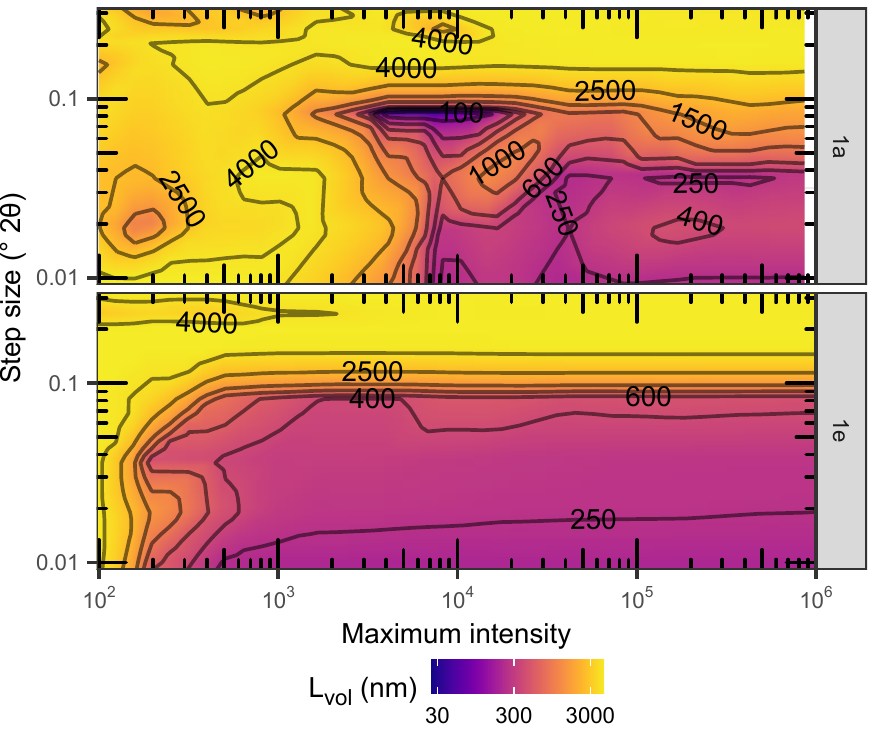}
\caption{Crystallite size for corundum in sample 3/150. It can be seen that they only agree for small step sizes and large maximum intensities, due to the ability to properly resolve the requisite peaks.}
\label{fig:SI_cryssize}
\end{figure}
%%%%

%%%%
\begin{figure}
\includegraphics{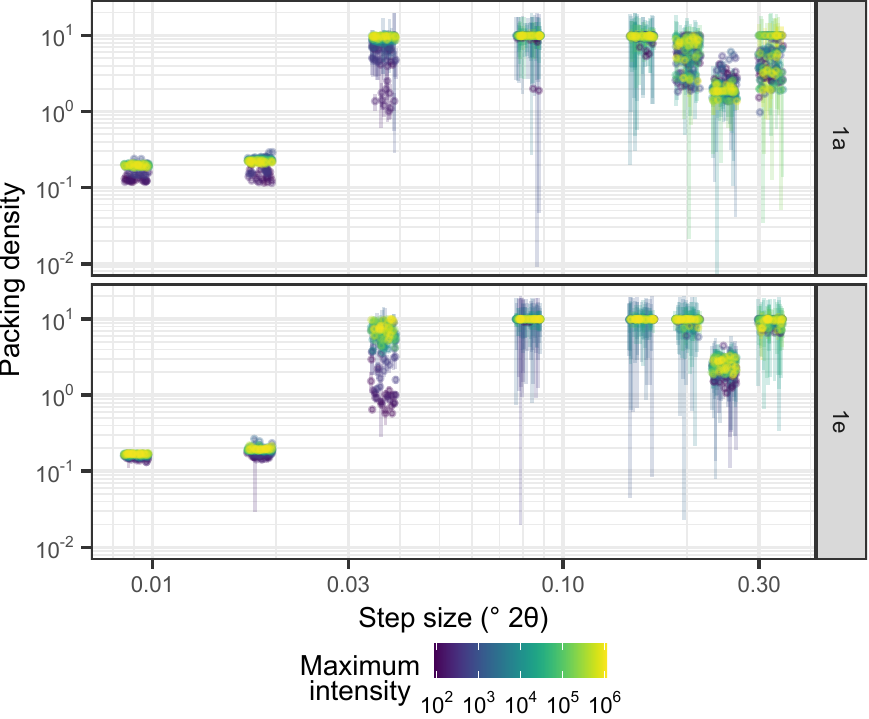}
\caption{Packing density for samples 1a and 1e for all intensities and HALs for refinement types 3 and 4. Error bars represent twice the standard uncertainty. The datapoints have been displaced slightly from their x-axis values for clarity.}
\label{fig:SI_pd}
\end{figure}
%%%%

%%%%
\begin{figure}
\includegraphics{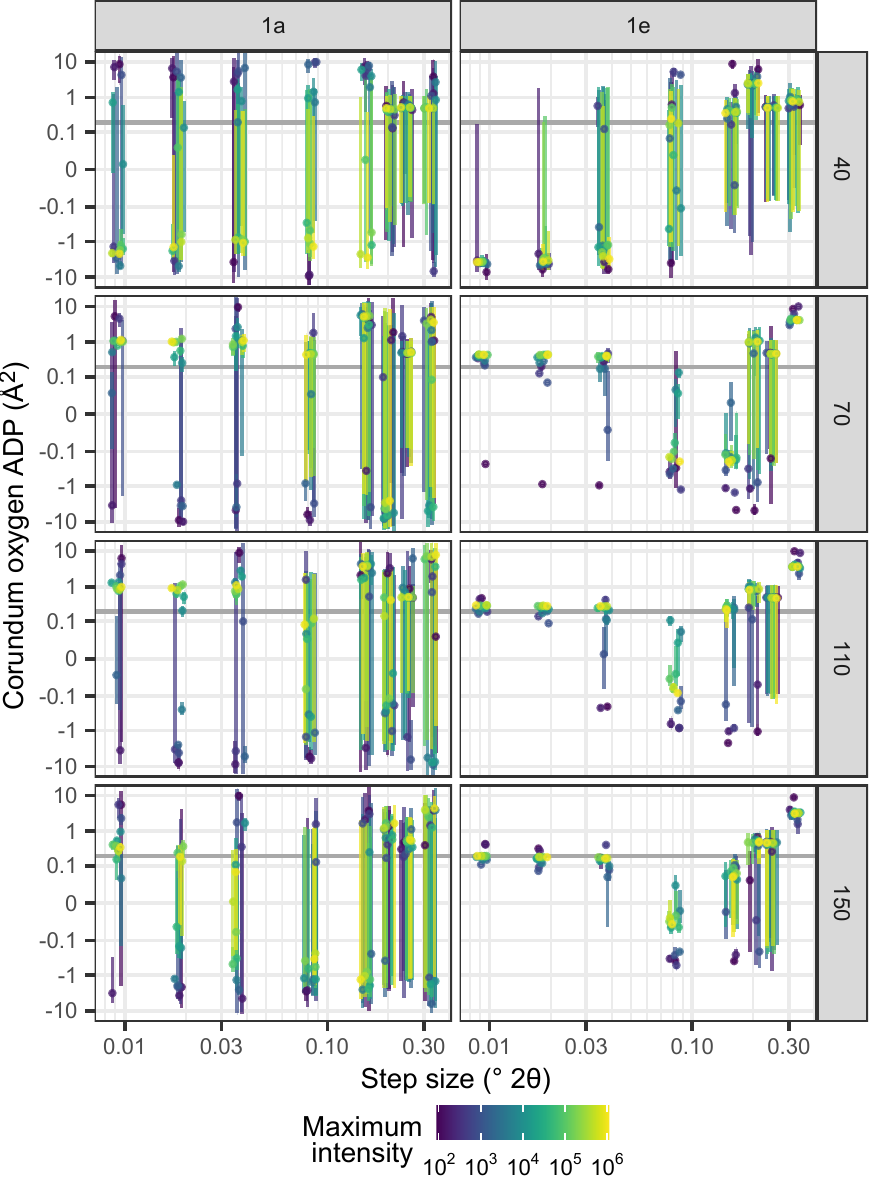}
\caption{ADP estimates for oxygen in corundum for  for samples 1a and 1e, refinement type 4, and the given HALs. The horizontal gray line represents the value given in Table~\ref{table:modelfinish}. The error bars represent twice the standard deviation. The standard uncertainty is only of significance for low intensity, small HAL, large stepsize patterns. Please note that the vertial axis is logarithmic. The datapoints have been displaced slightly from their x-axis values for clarity.}
\label{fig:SI_tp}
\end{figure}
%%%%

%%%%
\begin{figure}
\includegraphics{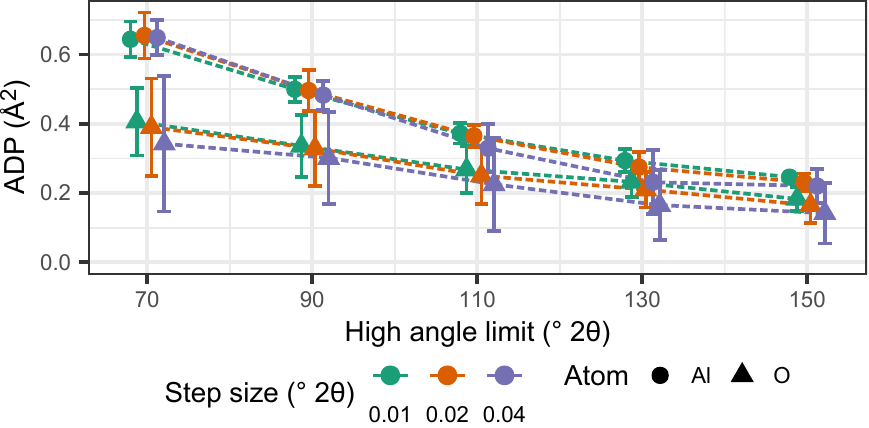}
\caption{ADPs for corundum in samples 1e/$>$~\num{1000}. Error bars represent twice the standard deviation of estimates averaged over all intensities. The datapoints have been displaced slightly from their x-axis values for clarity.}
\label{fig:SI_tp2}
\end{figure}
%%%%

%%%%%
%\begin{figure}
%\includegraphics{figures/SI/zesd_02_SI.pdf}
%\caption{Correlation between zero error and specimen displacement for samples \numrange{0.01}{0.08}/\numrange{90}{150}. The range of each refinement type is circled. 1a is the upper line, 1e is the lower line. The error bars represent twice the standard deviation of each estimate, with the standard uncertainty being insignificant on this scale. }
%\label{fig:SI_zesd}
%\end{figure}
%%%%%

%%%%
\begin{figure}
\includegraphics{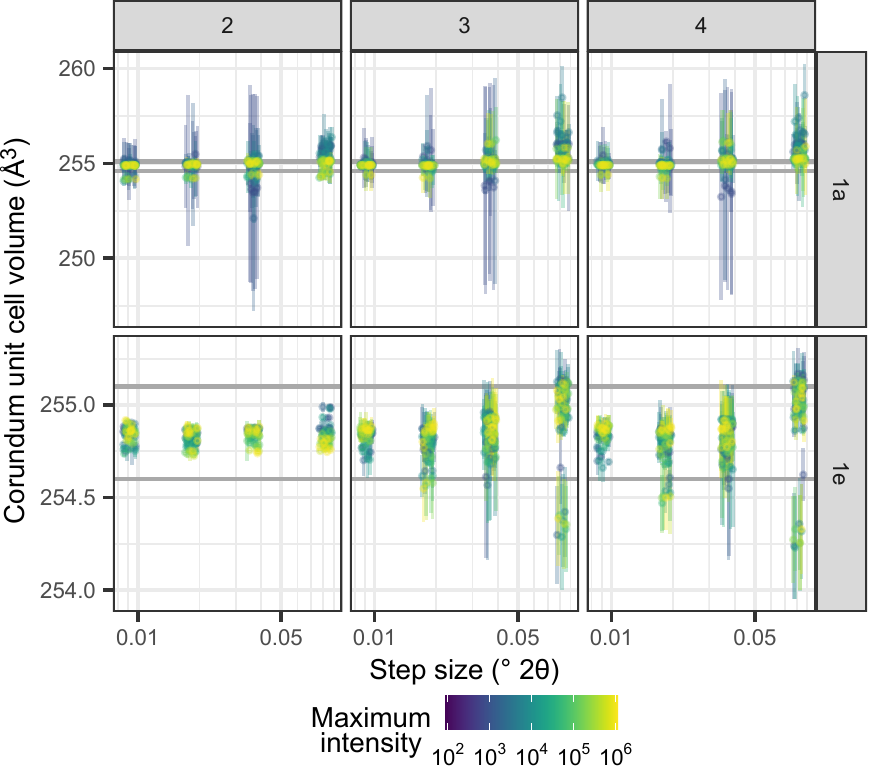}
\caption{Spread of calculated unit cell volume for corundum in by refinement type for samples 1a and 1e for all intensities $>$ 500 and all HALs. Error bars represent twice the standard deviation of the estimates. Please note that the vertical axis in the two rows are different; guidelines have been drawn in both rows to indicate an identical vertical range in both rows. The datapoints have been displaced slightly from their x-axis values for clarity.}
\label{fig:SI_vol}
\end{figure}
%%%%

%%%%
\begin{figure}
\includegraphics{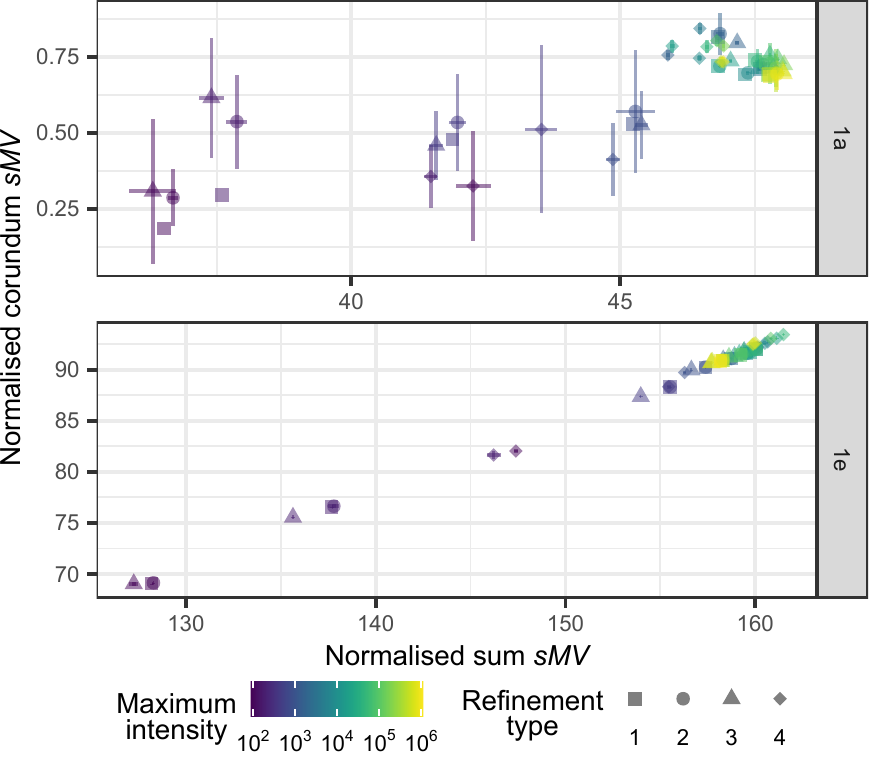}
\caption{Comparison of the sMVs for corundum and sum for samples 0.02/110, normalised by maximum intensity. Similar behaviour is exhibited for all refinements. The error bars represent twice the standard deviation.}
\label{fig:SI_smv}
\end{figure}
%%%%

%%%%%
%\begin{figure}
%\includegraphics{figures/SI/smvcor_01_SI.pdf}
%\caption{Correlation of corundum scale factors and unit cell volumes for all 200 refinements of sample 1e/3/110/0.02/\num{20000}. The Pearson correlation coefficient is 0.85; similar behaviour is seen for all refinements. Correlation reduces with increasing step size and decreasing HAL. Error bars represent twice the esd.}
%\label{fig:SI_smvcor}
%\end{figure}
%%%%%

%%%%
%\onecolumn
\begin{figure}
\includegraphics{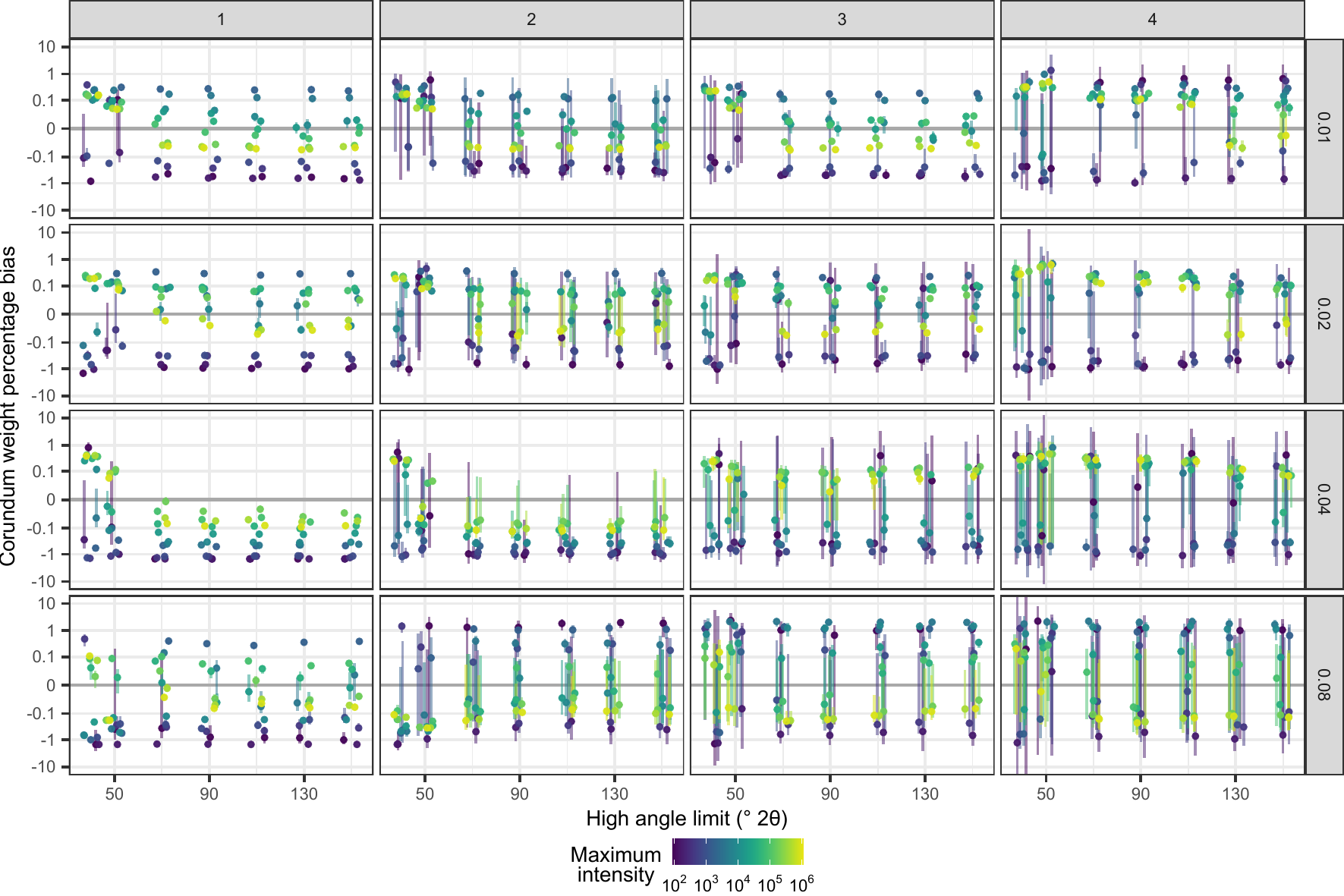}
\caption{Absolute corundum weight percentage bias in sample 1a separated by refinement type and stepsize. Error bars represent twice the combined standard deviation and uncertainty. Please note that the vertical axis is logarithmic.  For step sizes $>$~\SI{0.08}{\degTTh}, the bias is similar to that of \SI{0.08}{\degTTh}, increasing to $\sim$~\numrange{20}{50} percentage points for \SI{0.32}{\degTTh}. The datapoints have been displaced slightly from their x-axis values for clarity.}
\label{fig:SI_wt1a}
\end{figure}
%\twocolumn
%%%%

%%%%
%\onecolumn
\begin{figure}
\includegraphics{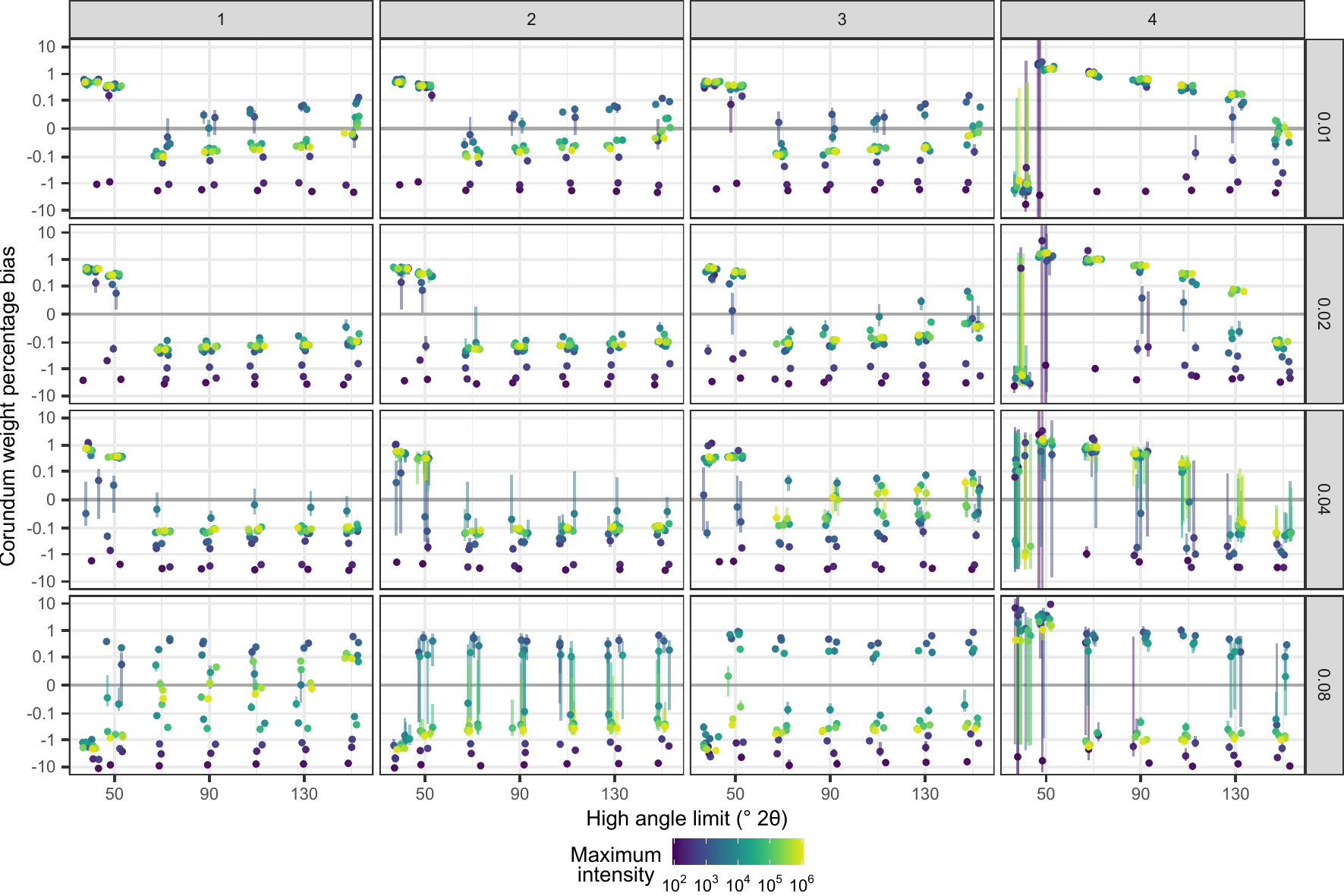}
\caption{Absolute corundum weight percentage bias in sample 1e separated by refinement type and stepsize. Error bars represent twice the combined standard deviation and uncertainty. Please note that the vertical axis is logarithmic. For step sizes $>$~\SI{0.08}{\degTTh}, the bias is similar to that of \SI{0.08}{\degTTh}, increasing to $\sim 20$ percentage points for \SI{0.32}{\degTTh}. The datapoints have been displaced slightly from their x-axis values for clarity.}
\label{fig:SI_wt1e}
\end{figure}
%\twocolumn
%%%%

%%%%
\begin{figure}
\includegraphics{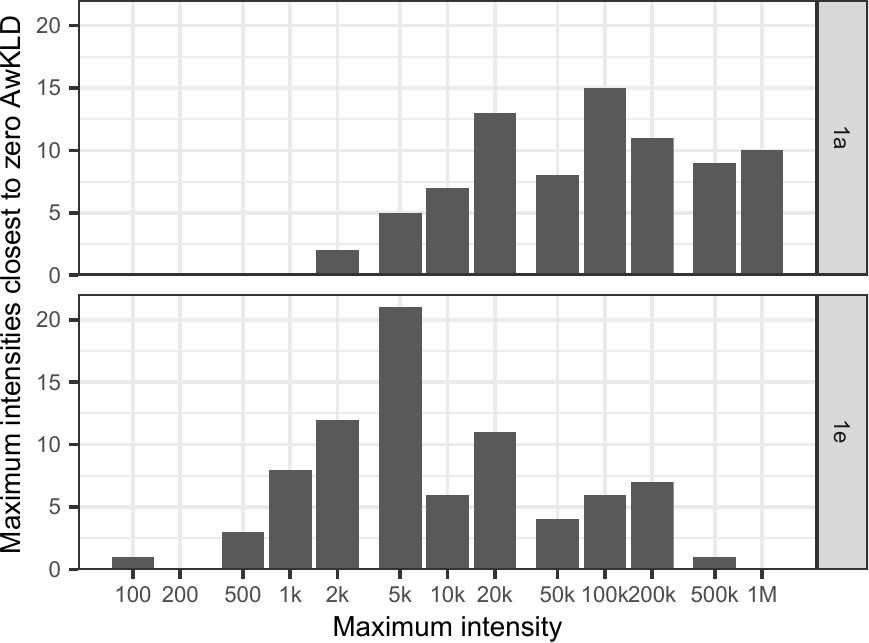}
\caption{Histograms showing the number of diffraction patterns for which that particular maximum intensity had the AwKLD value closest to zero, for samples 1a and 1e. The count is taken over samples $\ge$~70/$\le$~0.08. It can clearly be seen that the sample with phases present in smaller amounts requires maximum intensities much greater than the sample with phases present at approximately equal amounts.}
\label{fig:SI_wKLD_maxint}
\end{figure}
%%%%

%%%%
\begin{figure}
\includegraphics{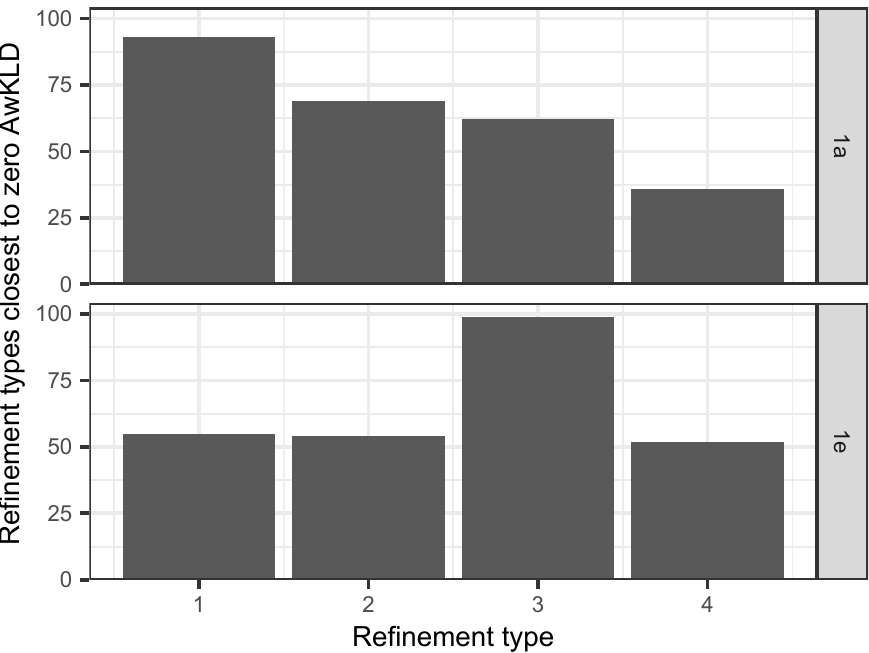}
\caption{Histograms showing the number of diffraction patterns for which that refinement type had the AwKLD value closest to zero, for samples 1a and 1e. The count is taken over samples $\ge$~70/$\le$~0.08. It can clearly be seen that the sample with phases present in smaller amounts requires a more constrained refinement than the sample with phases present at approximately equal amounts, where even this benefits from not refining ADPs.}
\label{fig:SI_wKLD_reftype}
\end{figure}
%%%%

%%%%
\begin{figure}
\includegraphics{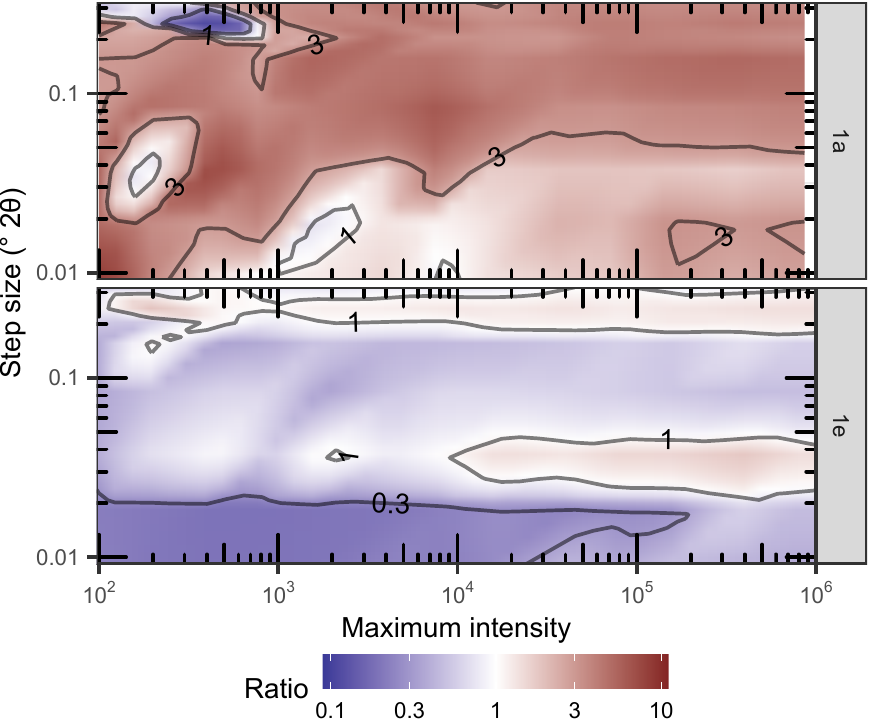}
\caption{The ratio, $\sigma_R/\sigma_T$, of the uncertainties of the weight fraction of corundum from sample 4/130 calculated from this study (Eqn~\ref{eqn:myerror}) and from \cite{Toraya-JAC-2000} (Eqn~\ref{eqn:torayaerror}), with $D=1$. It is clear that the Toraya method behaves substantially differently between evenly and unevenly distributed intensities.}
\label{fig:SI_torayaratio}
\end{figure}
%%%%
%\caption{The ratio, $\sigma_R/\sigma_T$, of the uncertainties of the weight fraction of corundum from sample 3/130 calculated from this study (Eqn~\ref{eqn:myerror}) and from \citeasnoun{Toraya-JAC-2000} (Eqn~\ref{eqn:torayaerror}), with $D=1$. It is clear that the Toraya method is systematically overestimating errors for evenly distributed intensities, but is inconsistent for uneven.}

%%%%
\begin{figure}
\includegraphics{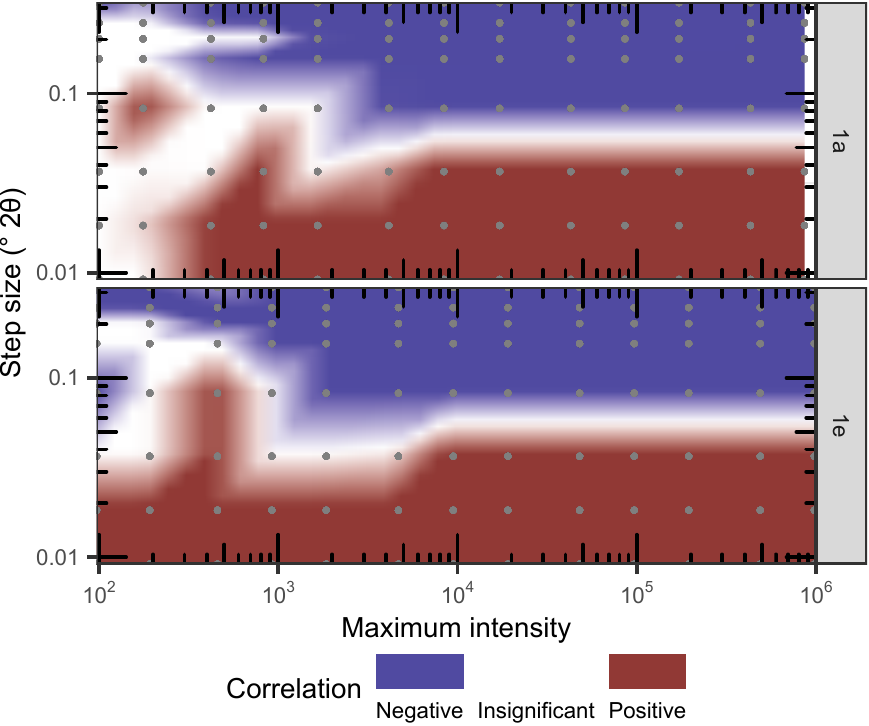}
\caption{Regions of negative, insignificant, and positive serial correlation for sample 4/130 according to Eqns~\ref{eqn:positive} -- \ref{eqn:Q}. The position of the data making the plot is given by the grey points. The overall behaviour of this plot is the same for all others, with large regions of positive and negative serial correlation present at the bottom and top of the plot, the long white bar moving down slightly, and the large white region becoming a little larger.}
\label{fig:SI_goodwDW}
\end{figure}
%%%%

\bibliography{row}